\documentclass{article}
\usepackage{latexsym}
\usepackage{amscd}
\usepackage{amsmath}
\newcounter{saveeqn}%
\newcommand{\alphaeqn}{\setcounter{saveeqn}{\value{equation}}%
\stepcounter{saveeqn}\setcounter{equation}{0}%
\renewcommand{\theequation}
        {\mbox{\arabic{saveeqn}-\alph{equation}}}}%
\newcommand{\reseteqn}{\setcounter{equation}{\value{saveeqn}}%
\renewcommand{\theequation}{\arabic{equation}}}%

\begin{document}
\title{Symmetries and conservation laws in histories-based generalized quantum mechanics}
\author{Tulsi Dass\thanks{Email: tulsi@iitk.ac.in}  and Yogesh Joglekar\thanks{Present address: Department of Physics, Indiana University, Bloomington, IN 47405} \footnote{email: yojoglek@indiana.edu } \\ Department of Physics \\ Indian Institute of Technology, Kanpur, India 208016}
\date{}
\maketitle


\begin{abstract}
Symmetries are defined in histories-based generalized quantum mechanics paying special attention to the class of history theories admitting quasitemporal structure (a generalization of the concept of `temporal sequences' of `events' using partial semigroups) and logic structure for `single time histories'. Symmetries are classified into orthochronous (those preserving the `temporal order' of `events') and non-orthochronous. A straightforward criterion for physical equivalence of histories is formulated in terms of orthochronous symmetries; this criterion covers various notions of physical equivalence of histories considered by Gell-Mann and Hartle as special cases. In familiar situations, a reciprocal relationship between traditional symmetries (Wigner symmetries in quantum mechanics and Borel-measurable transformations of phase space in classical mechanics) and symmetries defined in this work is established. In a restricted class of theories, a definition of conservation law is given in the history language which agrees with the standard ones in familiar situations; in a smaller subclass of theories, a Noether type theorem (implying a connection between continuous symmetries of dynamics and conservation laws) is proved.
\end{abstract}


\section{INTRODUCTION}
\label{one}

Consistent histories approach to quantum mechanics [1-7] has been employed recently for the treatment of some fundamental questions in physics. In the current histories-based quantum theory, by history of a system S one generally means a time-ordered sequence of `events' of the form
\begin{equation}
\label{1.1}  		
\alpha = (\alpha_{t_1}, \alpha_{t_2}, \ldots, \alpha_{t_n}); \hspace{1in} t_1 < t_2 < \ldots < t_n 
\end{equation}				
(where $\alpha_{t_i}$ are the Schr\"{o}dinger picture projection operators) given the system to be in the state represented by the density operator $\rho(t_0)$ at time $t_0 < t_1$. In traditional quantum mechanics (assuming, as usual, that the projections in eq.(\ref{1.1}) represent measurements by external observers) the probability of the history (\ref{1.1}) is given by 
\begin{equation}
\label{1.2}	     
P(\alpha) = \mbox{Tr}\left[ C_\alpha \rho(t_0) C_\alpha^\dagger \right]
\end{equation}				
where
\begin{equation}
\label{1.3}		
C_\alpha = \alpha_{t_n} U(t_n,t_{n-1}) \alpha_{t_{n-1}} \ldots \alpha_{t_2}U(t_2,t_1) \alpha_{t_1} U(t_1,t_0),
\end{equation}				
and $U(t,t') = \exp\left[-iH(t-t')/\hbar\right] $ is the evolution operator.

In the interpretive scheme of Griffiths and Omnes [1-3], the Hilbert-space based mathematical formalism is retained, the reduction postulate is discarded and eq.(\ref{1.2}) is proposed to be interpreted as the probability for the history (\ref{1.1}) for a \emph{closed} system S (no external observers). As all the probabilities employed are classical, the probability assignment can be made only for histories satisfying appropriate `consistency conditions' (or `decoherence conditions') ensuring the absence of quantum mechanical interference in the relevant family of histories. For certain pairs of histories $\alpha, \beta$ with $\alpha$ as in eq.(\ref{1.1}) and $ \beta = ( \beta_{t_1}, \ldots, \beta_{t_m})$ with the same initial state $\rho(t_0)$, the decoherence condition takes the form 
\begin{equation}
\label{1.4}		
\mbox{Re}[d(\alpha,\beta)] = 0,
\end{equation}				
where the so-called \emph{decoherence functional} $d(\alpha,\beta)$ is given by 
\begin{equation}
\label{1.5}
d(\alpha,\beta) = \mbox{Tr} [C_\alpha \rho(t_0) C_\beta^\dagger].
\end{equation}				
Note that
\begin{equation}
\label{1.6}			
P(\alpha) = d(\alpha,\alpha). 
\end{equation}				

In a formalism with histories as the basic objects, the time sequence $(t_1,\ldots, t_n)$ employed in the description of histories like (\ref{1.1}) serve only for book-keeping; the properties of time $t$ as a real variable are not used. The mathematical structure which correctly describes the book-keeping and also serves to make provision for generalization of the concept of time in histories-based theories is that of a partial semigroup [8].

A \emph{partial semigroup} (psg) is a nonempty set $\mathcal{K}$ (whose elements will be denoted as $s, t, u, \ldots$) in which a binary operation $\circ$ between certain pairs of elements is defined such that $(s\circ t)\circ u = s\circ(t\circ u)$ whenever both sides are well-defined. A homomorphism of a psg $\mathcal{K}$ into another psg $\mathcal{K}'$ is a mapping $\sigma : \mathcal{K} \rightarrow \mathcal{K}'$ such that, for all  $ s, t \in \mathcal{K}$ with $s\circ t$ defined, $\sigma(s)\circ\sigma(t)$ is also defined and 
\begin{equation}
\label{1.7}		
\sigma(s \circ t) = \sigma(s)\circ\sigma(t).
\end{equation}				
If $\sigma$ is invertible, it is called an isomorphism (automorphism if $\mathcal{K}'$= $\mathcal{K}$). The terms anti-homomorphism, anti-isomorphism and antiautomorphism are similarly defined with the order of terms on the right in eq.(\ref{1.7}) reversed.

The partial semigroups involved in the book-keeping of histories of the form (\ref{1.1}) are $\mathcal{K}_1$ and $\mathcal{K}_2$ defined as follows. We have
\begin{displaymath}			
\mathcal{K}_1 = \{\mbox{finite ordered subsets of \emph{R}}\}.
\end{displaymath}			          
A general element $t \in \mathcal{K}_1$ is of the form 
\begin{equation}
\label{1.8}		
t=\{t_1,t_2, \ldots, t_n\}; \hspace{1in} t_1<t_2<\ldots < t_n.
\end{equation}				
If $s =\{s_1,s_2,\ldots,s_m\}\in\mathcal{K}_1$ such that $s_m < t_1$, then $s\circ t$ is defined and 
\begin{equation}
\label{1.9}		
s\circ t = \{s_1, s_2, \ldots, s_m, t_1,t_2,\ldots,t_n\}.
\end{equation}				
We shall adopt the convention 
\begin{equation}
\label{1.10}		
\{t_1\}\circ\{t_1\} = \{t_1\}.
\end{equation}				
With this convention, we have $s\circ t$ defined for $s_m\leq t_1$. Note that elements of $\mathcal{K}_1$ admit irreducible decomposition of the form
\begin{equation}
\label{1.11}		
t = \{t_1\}\circ \{t_2\}\circ \cdots \circ \{t_n\}.
\end{equation}				
Elements $\{t_i\}$ which cannot be further decomposed are called \emph{nuclear}.

The other psg $\mathcal{K}_2$ consists of histories of the form (\ref{1.1}) as its elements. For 
\begin{displaymath}		   	
\alpha = \{\alpha_{s_1}, \alpha_{s_2}, \ldots, \alpha_{s_m}\},\hspace{0.5in} \beta = \{\beta_{t_1},\beta_{t_2},\ldots,\beta_{t_n}\}
\end{displaymath}			
with $s_m< t_1$, $\alpha\circ\beta$ is defined and is given by
\begin{equation}
\label{1.12}		
\alpha\circ\beta =\{\alpha_{s_1},\alpha_{s_2},\ldots,\alpha_{s_m},\beta_{t_1},\beta_{t_2}, \ldots, \beta_{t_n}\}.		
\end{equation}				
There is a homomorphism $\sigma$ from $\mathcal{K}_2$ onto $\mathcal{K}_1$ given by
\begin{eqnarray}
\label{1.13}		
\sigma(\alpha) = s, & \sigma(\beta)= t, & \sigma(\alpha\circ\beta) = s\circ t.
\end{eqnarray}				
The triple $(\mathcal{K}_2,\mathcal{K}_1, \sigma)$ defines what Isham calls a \emph{quasitemporal} structure (a pair of psg's with a homomorphism of one onto the other). Note that, given a single time element $\{t_1\} \in \mathcal{K}_1$, the space $\left(\mathcal{K}_2\right)_{t_1} = \sigma^{-1}\{t_1\}$ is the set $\mathcal{P}(\mathcal{H})$ of projection operators in the quantum mechanical Hilbert space $\mathcal{H}$ of the system; in the framework of quantum logic [9-10], these projection operators represent single-time propositions. The space $\mathcal{P}(\mathcal{H})$ constitutes a \emph{logic} in the sense of Varadarajan [10].

The concept of quasitemporal structure appears to be the appropriate generalization of the idea of histories as temporal sequences of `events'. With suitably chosen psg's (employing light cones etc.) this concept serves to provide a framework general enough to accommodate history versions of quantum field theories in curved space-times [8].

Taking clue from the traditional proposition calculus [9-10] where single time propositions are taken as the basic entities, Isham [8] made another important suggestion: a formalism in which histories are to be the basic objects must treat them as (multitime or more general) propositions. He evolved a scheme of `quasitemporal theories' in which the basic objects were a triple $(\mathcal{U},\mathcal{T},\sigma)$ defining a quasitemporal structure. The space $\mathcal{U}$ was called the `space of history filters' and was assumed to be (besides being a psg) a meet semilattice with the operations of partial order $\leq$ (coarse graining) and a meet operation $\wedge$ (simultaneous realization of two histories). The space $\mathcal{T}$ was called the `space of temporal supports'. To accommodate the operation of negation of a history, the space $\mathcal{U}$ was proposed to be embedded in a larger space $\Omega$ (denoted as $\mathcal{UP}$ in [8] and [11]), called the `space of history propositions'. This larger space was envisaged as having a lattice structure. Decoherence conditions and probabilities of decoherent histories were supposed to be defined in terms of decoherence functionals which were complex valued functions defined on pairs of history propositions satisfying the standard four conditions [4,8,11] of hermiticity, positivity, bilinearity and normalization. (See eqs.(\ref{2.10a}-d) below.)  
						
A more general scheme was later proposed by Isham and Linden [11] in which the basic object was a pair of spaces $(\Omega,\mathcal{D})$ where $\Omega$ (the space of history propositions) was assumed to be an orthoalgebra incorporating a partial order $\leq$ (coarse graining), a disjointness relation $\bot$ (mutual exclusion), a join operation $\oplus$ (`or' operation for mutually exclusive propositions) and a few other features. The space $\mathcal{D}$ was the space of decoherence functionals satisfying the above mentioned properties. The quasitemporal theories [8] are a subclass of this general class of theories.

In a recent paper [12] we have presented axiomatic development of dynamics of  systems in the framework of histories which contains the history versions of classical and traditional quantum mechanics as special cases. We considered theories which admit quasitemporal structure $(\mathcal{U},\mathcal{T},\sigma)$ in the sense explained above. The spaces $\mathcal{U}_\tau = \sigma^{-1}(\{\tau\})$ for nuclear elements $\{\tau\}\in\mathcal{T}$ (the spaces of single time propositions) were assumed to have the structure of a logic [10]. Isomorphism of $\mathcal{U}_\tau$'s (as logics) at different $\tau$'s was not assumed. Using the logic structure of $\mathcal{U}_\tau$'s, a larger space $\Omega$ - the space of history propositions (`inhomogeneous histories') was explicitly constructed and shown to be an orthoalgebra as envisaged in the scheme of Isham and Linden [11]; its subspace $\tilde{\mathcal{U}}$ representing `homogeneous histories' (which is obtained from $\mathcal{U}$ after removing some redundancies) was shown to be a meet semilattice as envisaged in the scheme of [8]. Decoherence functionals satisfying the usual conditions are supposed to be constructed in terms of an initial sate and an evolution map (see eq.(\ref{1.5})). Explicit expressions for decoherence functionals were given for the Hilbert space based theories with $\mathcal{U}_\tau = \mathcal{P}(\mathcal{H}_\tau)$ (the lattice of projection operators in a separable Hilbert space $\mathcal{H}_\tau$) and for classical mechanics. 

The present work is devoted to a systematic treatment of symmetries and conservation laws in histories-based theories. We have chosen the formalism of [12] for a detailed treatment of symmetries; this is because the mathematical-physical structure of this formalism facilitates treatment of some detailed features of symmetry operations in history theories. Following the general idea [13] of defining symmetries as structure-preserving invertible mappings in appropriate mathematical framework, we define a symmetry as a triple $\Phi = (\Phi_1,\Phi_2,\Phi_3)$ of invertible mappings ($\Phi_1:\mathcal{T}\rightarrow \mathcal{T},\Phi_2:\mathcal{U}\rightarrow \mathcal{U}$ and $\Phi_3:\mathcal{D}\rightarrow \mathcal{D}$ where $\mathcal{D}$ is the space of decoherence functionals) preserving the quasitemporal structure, the logic structure of single `time' propositions, the decoherence condition and the probabilities of histories (see section \ref{three} for details). A natural classification of symmetries as \emph{orthochronous} and \emph{non-orthochronous} appears. In the case of (history version of) traditional quantum mechanics and classical mechanics, a reciprocal relationship is established between symmetries defined here and symmetries of the formalism of traditional sort - Wigner symmetries in quantum mechanics and Borel measurable transformations of the phase space in classical mechanics (the latter are natural structure preserving transformations in classical mechanics in the framework of logics).
								
Gell-Mann and Hartle [14] have considered several notions of physical equivalence of histories in the quantum mechanics of closed systems. Using the concept of orthochronous symmetries we obtain an economic formulation of the concept of physical equivalence of histories which covers all these notions as special cases (section \ref{three}). From our definition of symmetry in the formalism of [12] it is clear how symmetries are to be defined in general quasitemporal theories [8] and more generally, in the Isham and Linden scheme [11]. A good example in this context is the work of Schreckenberg [15] who has considered symmetries in the subclass of Isham Linden type theories treated in [16]. This work is also briefly described in section \ref{three}. 

It is of some interest to treat conservation laws in the framework of histories. As might be intuitively anticipated, conservation laws can be defined only if the spaces of single `time' propositions (the $\mathcal{U}_\tau$'s mentioned above) at various instants of `time' are mutually isomorphic. We first give a straightforward definition of conservation law in terms of temporal evolution described by mappings between pairs of $\mathcal{U}_\tau$'s. It is then translated into a (supposedly equivalent) definition in terms of decoherence functionals. The equivalence of the two definitions is verified in (history versions of) traditional quantum mechanics and classical mechanics. We do not consider the deeper question of the conservation of an observable in histories involving `events' relating to other observables as well and the constraints implied by such conservation. Such questions have been considered by Hartle {\it et al}. [17] in the context of the Hilbert space quantum mechanics (of closed systems). That work, which employs a definition of conservation law somewhat different from ours, is very briefly described in section \ref{four}.

In any scheme of mechanics, one expects a general connection between continuous symmetries of dynamics and conservation laws. The most famous result of this type is the Noether's theorem [18,19] in Lagrangian dynamics. In our formalism, the twin requirements of an explicit expression for the decoherence functional and interpretation of the infinitesimal generators of a symmetry in terms of observables appears to restrict the possibility of proving such a theorem only for the Hilbert space based theories in which the Hilbert spaces $\mathcal{H}_\tau$ corresponding to various nuclear $\tau$ are naturally isomorphic. A theorem showing that, in such theories, a continuous symmetry implies a conservation law is proved in section \ref{seven}.

The rest of the paper is organized as follows. In the next section, we recapitulate the main developments in [12] collecting equations needed for reference in later sections. Section \ref{three} is devoted to the treatment of symmetries in the formalism of [12]; this section also includes the items relating to [14] and [15,16] mentioned above, a brief account of the treatment of symmetries in the work of Houtappel, Van Dam and Wigner (HVW) [20] which was the first work to give a treatment of symmetries in a framework of histories-like objects. The next two sections are devoted to the comparison of traditional and present treatment of symmetries in nonrelativistic quantum mechanics and exhibiting the two-way connection mentioned above. Sections \ref{six} and \ref{seven} are devoted, respectively, to the treatment of conservation laws and a Noether type theorem as mentioned above. The last section contains some concluding remarks.			


\section{ HISTORIES-BASED GENERALIZED QUANTUM MECHANICS}
\label{two}

In this section, we shall recapitulate some essential points from [12]. First we recall a few more points relating to the partial semigroups.


\subsection{More about partial semigroups}
A \emph{unit element} in a psg $\mathcal{K}$ is an element $e$ such that $e\circ s = s\circ e = s$ for all $s \in \mathcal{K}$. An \emph{absorbing element} in $\mathcal{K}$ is an element $a$ such that for all $s\in \mathcal{K}$, $a\circ s = s\circ a = a $. A psg may or may not have a unit and/or absorbing element; when either of them exists, it is unique. In a psg, elements other than the unit and the absorbing elements will be called \emph{typical}. If a psg $\mathcal{K}$ has a unit element $e$ and/or an absorbing element $a$ and if there is a homomorphism $\sigma$ from $\mathcal{K}$ onto $\mathcal{K}'$, then  $\mathcal{K}'$ must correspondingly have a unit element $e'$ and an absorbing element $a'$ such that 
\begin{equation}
\label{2.1}
\sigma(e) = e', \hspace{.3in} \sigma(a) = a'.
\end{equation}				

A psg $\mathcal{K}$ is called \emph{directed} if, for any two different typical elements $s, t$ in $\mathcal{K}$, when $s\circ t$ is defined then $t\circ s$ is not defined. The psg's $\mathcal{K}_1$ and $\mathcal{K}_2$ introduced in section \ref{one} are directed. In [12], it was suggested that, in a theory with a quasitemporal structure, the concept of direction of flow of `time' could be introduced by taking the two psg's to be directed.

The concept of `point of time' gets replaced in the psg setting by that of a nuclear element. The nuclear elements of $\mathcal{K}_1$ are, indeed, points of time. (See eq.(\ref{1.11}).) In $\mathcal{K}_2$, the nuclear elements are of the form $\{\alpha_t\}$ representing single time history propositions. The set of nuclear elements in a psg $\mathcal{K}$ will be denoted as $\mathcal{N}(\mathcal{K})$. Clearly $\mathcal{N}(\mathcal{K}_1)=R $, the set of real numbers.

In [12], the concept of \emph{special} psg's was introduced. These are the psg's whose typical elements admit semi-infinite irreducible decompositions and which incorporate conventions of the form (\ref{1.10}). The psg's $\mathcal{K}_1$ and $\mathcal{K}_2$ are trivial examples of the special psg's. A nontrivial example [11] is the psg $\mathcal{K}_3$ whose elements are ordered subsets of $R$ which are at most countably semi-infinite (with elements of the form $ s=\{s_1,\ldots,s_n\}, s' = \{s_1, s_2,\ldots\}, s''=\{\ldots,s_{-1}, s_0\} $) with composition rule a straightforward extension of that of $\mathcal{K}_1$.


\subsection{The augmented temporal logic formalism}
 The axiomatic development of generalized quantum mechanics of closed systems (which contains history versions of classical and quantum mechanics as special cases) presented in [12] is structured around five axioms $A_1,\ldots,A_5$.

\begin{enumerate}
\item[]{\bfseries \emph{$A_1$: (Quasitemporal Structure Axiom)}}: 
 Associated with every dynamical system is a history system $(\mathcal{U},\mathcal{T},\sigma)$ defining a quasitemporal structure as in Isham's formalism [8]. The psg's $\mathcal{U}$ and $\mathcal{T}$ are assumed to be special and satisfy the relation 
\begin{equation}
\label{2.2}		
\sigma\left[ \mathcal{N}(\mathcal{U})\right] = \mathcal{N}(\mathcal{T}).
\end{equation}				
\end{enumerate}

Elements of $\mathcal{U}$ (history filters) will be denoted as $\alpha,\beta,\ldots$ and those of $\mathcal{T}$ (temporal supports) as $\tau,\tau',\ldots$. If $\alpha\circ\beta$ and $\tau\circ\tau'$ are defined, we write $\alpha\lhd \beta$ ($\alpha$ precedes $\beta$) and $\tau\lhd\tau'$ ($\tau$ precedes $\tau'$).

\begin{enumerate}
\item[] {\bfseries \emph{$A_2$: (Causality Axiom)}}: If $\alpha,\beta,\ldots,\gamma\in\mathcal{N}(\mathcal{U})$ are such that $\alpha\lhd\beta\lhd \ldots\lhd\gamma$ with $\sigma(\alpha) = \sigma(\gamma)$ then we must have $\alpha = \beta = \ldots = \gamma$.
\end{enumerate}

In essence this axiom forbids histories corresponding to `closed time loops'. From these two axioms we can prove that the psg's $\mathcal{U}$ and $\mathcal{T}$ are directed (and some other useful results [12].)

\begin{enumerate}
\item[] {\bfseries\emph{$A_3$: (Logic Structure Axiom)}}: Every space $\mathcal{U}_\tau = \sigma^{-1}(\tau)$ for $\tau\in \mathcal{N}(\mathcal{T})$ has the structure of a logic as defined in [10].
\end{enumerate}


We do not assume isomorphism of $\mathcal{U}_\tau$'s (as logics) for different $\tau$'s. Every $\mathcal{U}_\tau$ has two distinguished elements $0_\tau$ (the null proposition) and $1_\tau$ (the unit proposition) such that $0_\tau\leq\alpha\leq 1_\tau$ for all $\alpha\in\mathcal{U}_\tau$. For a typical  $\alpha \in \mathcal{U}_\tau$, we define $supp(\alpha)$ (called the temporal support of $\alpha$) as the unique collection of elements of $\mathcal{N}(\mathcal{T})$ appearing in the irreducible decomposition of $\sigma(\alpha)$. For any $\tau\in supp(\alpha)$, the nuclear element in the irreducible decomposition of $\alpha$ projecting onto $\tau$ under $\sigma$ is denoted as $\alpha_\tau$.

The possible presence of $0_\tau$'s and $1_\tau$'s in irreducible decompositions causes some redundancy which needs to be removed. We call an $\alpha \in \mathcal{U}$ a \emph{null history filter} if $\alpha_\tau=0_\tau$ for at least one $\tau \in supp(\alpha)$; we call it a \emph{unit history filter} if $\alpha_\tau = 1_\tau$ for all $\tau\in supp(\alpha)$. All the null history filters are physically equivalent (they represent absurd histories) as are all unit history filters. We remove the redundancy by introducing an equivalence relation in $\mathcal{U}$ such that all null history filters are treated as equivalent and so are all non-null filters with the same reduced form (the form obtained by deleting the redundant $1_\tau$'s in the irreducible decomposition). We denote the equivalence class of $\alpha\in\mathcal{U}$ by $\tilde{\alpha}$. The set $\tilde{\mathcal{U}}$ of the equivalence classes in $\mathcal{U}$ inherits a psg structure from $\mathcal{U}$. In this psg, the equivalence class $\tilde{0}$ of all null history filters acts as an absorbing element and the equivalence class $\tilde{1}$ of all unit history filters acts as the unit element. 
								
A psg $\tilde{\mathcal{T}}$ is constructed from $\mathcal{T}$ in a similar fashion. The typical elements of $\tilde{\mathcal{T}}$ and the homomorphism $\tilde{\sigma}:\tilde{\mathcal{U}} \rightarrow \tilde{\mathcal{U}}$ (restricted to typical elements of $\tilde{\mathcal{U}}$) is obtained by defining $\tilde{\tau} = \tilde{\sigma}(\tilde{\alpha})$ to be the reduced object obtained from $\tau = \sigma(\alpha)$ (for any representative $\alpha$ of $\tilde{\alpha}$) by deleting under $\sigma$ the images of the redundant $1_\tau$'s in the irreducible decomposition of $\alpha$. The unit element $\tilde{e}$ and the absorbing element $\tilde{a}$ are defined as
\begin{equation}
\label{2.3}	  	
\tilde{e} = \phi\hspace{.5cm}(\mbox{the empty subset of }\mathcal{N}(\mathcal{T})), \hspace{0.3in} \tilde{a} =  \mathcal{N}(\mathcal{T})	
\end{equation}				
and the homomorphism $\tilde{\sigma}$ is extended to include the relations 
\begin{equation}
\label{2.4}		
\tilde{\sigma}(\tilde{1}) = \tilde{e}, \hspace{0.3in} \tilde{\sigma}
(\tilde{0}) = \tilde{a} .
\end{equation}				
Defining, for any $\tilde{\alpha} \in \tilde{\mathcal{U}}, supp(\tilde{\alpha}) = \tilde{\sigma}(\tilde{\alpha})$ (considered as a subset of $\mathcal{N}(\tilde{\mathcal{T}}) = \mathcal{N}(\mathcal{T}) $), we have
\begin{equation}
\label{2.5}		
supp(\tilde{1}) = \phi, \hspace{1in} supp(\tilde{0}) = \mathcal{N}(\mathcal{T}).
\end{equation}				
 
Using irreducible decompositions and the logic structure on $\mathcal{U}_\tau$'s one can define (in an intuitively suggestive manner) partial order ($\leq$), disjointness ($\bot$), disjoint join operation ($\oplus$) and meet operation ($\wedge$) in $\tilde{\mathcal{U}}$. The structure $( \tilde{\mathcal{U}}, \leq, \wedge)$ was shown in [12] to be a meet semilattice as envisaged in Isham's scheme [8]. It is the triple $ (\tilde{\mathcal{U}}, \tilde{\mathcal{T}}, \tilde{\sigma})$ (and not the original $(\mathcal{U}, \mathcal{T},\sigma)$) which corresponds to the triple in [8].

The embedding of  $\tilde{\mathcal{U}}$ in a larger space $\Omega$ of history propositions envisaged in Isham's scheme is realized concretely in the present formalism by defining $\Omega$ to be the space of at most countable collections of mutually orthogonal elements of  $\tilde\mathcal{U}$ such that the union of temporal supports of any finite subcollection of them is orientable. (A subset $A$ of a psg $\mathcal{K}$ is said to be \emph{orientable} if it is at most countable and if there exists an ordering of elements of $A$ such that composition of every pair of consecutive elements is defined.) We denote elements of $\Omega$ as $\underline{\alpha},\underline{\beta},\ldots$. The space $\Omega$ has a null element $\underline{0}=\{\tilde{0}\}$ and a unit element $\underline{1}=\{\tilde{1}\}$. Elements of $\Omega$ other than $\underline{0}$ and $\underline{1}$ are called \emph{generic}.

One can define (again, in an intuitively suggestive manner), in $\Omega$, the operations of partial order ($\leq$), disjointness ($\bot$), disjoint join operation ($\oplus$) and show [12] that, with these operations $\Omega$ is an orthoalgebra (as envisaged in the scheme of Isham and Linden [11]). A general element $\underline{\alpha}=\{\tilde{\alpha}^{(1)},\tilde{\alpha}^{(2)},\ldots\}$ of $\Omega$ can also be represented as 
\begin{equation}
\label{2.6}		
\underline{\alpha} = \{\tilde{\alpha}^{(1)}\}\oplus\{\tilde{\alpha}^{(2)}\}
\oplus\cdots
\end{equation}				
A collection $\{\underline{\alpha},\underline{\beta},\ldots\}$ of mutually disjoint elements of $\Omega$ is said to be \emph{complete} (or \emph{exhaustive}) if
\begin{equation}
\label{2.7}		
\underline{\alpha}\oplus\underline{\beta}\oplus\cdots = \underline{1} .
\end{equation}				

The logic structure of $\mathcal{U}_\tau$'s permits us to introduce a space $\mathcal{S}(\tau)$ of states and a space $\mathcal{O}(\tau)$ of observables at `time' $\tau$ as in traditional proposition calculus. A state at `time' $\tau\in \mathcal{N}(\mathcal{T})$ is a generalized probability on $\mathcal{U}_\tau$, \emph{i.e.} a map $p_\tau:\mathcal{U}_\tau\rightarrow R$ such that (i) $ 0\leq p_\tau(\alpha)\leq 1$ for all $\alpha\in\mathcal{U}_\tau$, (ii) $ p_\tau (0_\tau) = 0$, $p_\tau (1_\tau)=1$ and (iii) it is countably additive in the sense that, given a sequence $\alpha_1,\alpha_2,\ldots$ of pairwise disjoint elements in $\mathcal{U}_\tau$, we have						
\begin{equation}
\label{2.8}		
p_\tau \left(\vee_{i} \alpha_i\right) = \sum_{i} p_\tau (\alpha_i)
\end{equation}				
where $\vee$ is the join operation in $\mathcal{U}_\tau$. An observable at `time' $\tau$ is a map $A_\tau: B(R) \rightarrow \mathcal{U}_\tau$ (where $B(R)$ is the $\sigma$-algebra of Borel subsets of $R$ - the smallest $\sigma$-algebra containing all the open intervals) such that (i) $A_\tau(\phi) = 0_\tau$,  $A_\tau(R) = 1_\tau$, (ii) given disjoint sets $E, F$ in $B(R)$, we have $A_\tau(E)$ and $A_\tau(F)$ disjoint in $\mathcal{U}_\tau$; (iii) if $E_1,E_2,\cdots$ is a sequence of pairwise disjoint sets in $B(R)$, we have
\begin{equation}
\label{2.9}		
A_\tau\left(\cup_k E_k \right)=\bigvee_k A_\tau(E_k).
\end{equation} 				


We can now state the last two axioms.
\begin{enumerate}
\item[]{\bfseries\emph{$A_4$:(Temporal Evolution)}} : The temporal evolution of the system with history system $(\mathcal{U},\mathcal{T},\sigma)$ is given, for each pair $\tau, \tau'\in\mathcal{N}(\mathcal{U})$ such that $\tau\lhd\tau'$, by a set of mappings $V(\tau',\tau)$ of $ \mathcal{U}_\tau$ onto $\mathcal{U}_\tau'$, which are  logic homomorphisms (not necessarily injective) and which satisfy the composition rule $V(\tau'',\tau')\cdot V(\tau',\tau) = V(\tau'',\tau)$ whenever $\tau\lhd\tau', \tau'\lhd\tau''$ and $\tau\lhd\tau''$.
\item[]{\bfseries\emph{$A_5$(a):(Decoherence Functionals)}}: Given a state $p_o \in \mathcal{S}(\tau_0)$ for some $\tau_0 \in \mathcal{N}(\mathcal{T})$ and a law of evolution $V(\tau',\tau)$, we have a decoherence functional $d = d_{p_0,V}$ which is a mapping from (a subset of) $\Omega\times\Omega$ into $C$ such that
\alphaeqn
\begin{alignat}{2}
\label{2.10a}		
\mbox{(i)}& \hspace{3.5cm}d(\underline{\alpha},\underline{\beta})^{*}=d(\underline{\beta},\underline{\alpha}) \hspace{1cm} & \mbox{(hermiticity)}\\ 
\mbox{(ii)}& \hspace{3.5cm}d(\underline{\alpha},\underline{\alpha})\geq 0\hspace{.9cm} & \mbox{(positivity)} 
\end{alignat}				
\begin{enumerate}
\item[(iii)] If $\underline{\alpha},\underline{\beta}, \cdots$ is an at most countable collection of pairwise disjoint elements of $\Omega$, we have   
\begin{equation}
\label{2.10c}	
\begin{split}					
d(\underline{\alpha}\oplus\underline{\beta}\oplus\cdots ,\underline{\gamma})=&\quad d(\underline{\alpha}, \underline{\gamma}) + d(\underline{\beta}, \underline{\gamma}) + \cdots  \\
& \mbox{(countable additivity)} \\
\end{split} 
\end{equation}					
\end{enumerate}					
\begin{equation}
\label{2.10d}			
\mbox{(iv)}\hspace{3.6cm}d(\underline{1},\underline{1}) = 1 \hspace{1cm}\mbox{(normalization)}.
\end{equation}					
\reseteqn
\end{enumerate}
The space of decoherence functionals will be denoted as $\mathcal{D}$.

A complete set $\mathcal{C}$ of history propositions is said to be (weakly) decoherent (or consistent) with respect to a decoherence functional $d$ if
\begin{equation}
\label{2.11}		
\mbox{Re}[d(\underline{\alpha},\underline{\beta})] = 0, \hspace{0.4in}\underline{\alpha}, \underline{\beta}\in \mathcal{C}, \hspace{.2in} \underline{\alpha} \neq \underline{\beta}.
\end{equation}				

\begin{enumerate}						
\item[]{\bfseries\emph{$A_5$(b):(Probability Interpretation)}}: The probability that a history $ \underline{\alpha}$ in a complete set $\mathcal{C}$ which is decoherent with respect to a decoherence functional $d$ is realized is given by
\begin{equation}
\label{2.12}		
P( \underline{\alpha}) = d( \underline{\alpha}, \underline{\alpha}).
\end{equation}
\end{enumerate}				
For $\underline{\alpha}, \underline{\beta}\in \mathcal{C}$, we have the classical probability sum rule:
\begin{eqnarray}
\label{2.13}		
P(\underline{\alpha}\oplus\underline{\beta}) & = & d(\underline{\alpha}\oplus\underline{\beta}, \underline{\alpha}\oplus\underline{\beta}) \nonumber \\
 & = & d(\underline{\alpha},\underline{\alpha}) + d(\underline{\beta},\underline{\beta}) + 2\,\mbox{Re } d(\underline{\alpha},\underline{\beta}) \nonumber \\
& = & P(\underline{\alpha}) + P(\underline{\beta}),
\end{eqnarray}				
and recalling eq.(\ref{2.7})
\begin{eqnarray}
\label{2.14}			
1 & = d(\underline{1},\underline{1}) & = d(\underline{\alpha}\oplus\underline{\beta}\oplus\cdots, \underline{\alpha}\oplus\underline{\beta}\oplus\cdots)  \nonumber \\
 & & = \sum_{\underline{\alpha}\in\mathcal{C}} P(\underline{\alpha})
\end{eqnarray}				
In [12], explicit expressions for $d_{p_0,V}$ were given for the Hilbert-space based theories and for classical mechanics. 


\section{SYMMETRIES IN HISTORIES-BASED GENERALIZED QUANTUM MECHANICS}
\label{three}

We shall start with a quick look at the HVW paper [20].


\subsection{Symmetries in the HVW approach}
Objects analogous to what are now called histories appear first in the work of Houtappel, Van Dam and Wigner (HVW) [20] who sought to present a general treatment of geometric invariance principles (\emph{i.e.} those invariance principles which correspond to transformations between equivalent reference frames) in classical and quantum mechanics in terms of the primitive elements of a physical theory. These primitive elements were taken to be the conditional probabilities $\Pi(A|B)$ where $A=(\alpha, r_\alpha;\beta,r_\beta,\cdots; \epsilon, r_\epsilon)$ represents a set of measurements $\alpha, \beta,\ldots, \epsilon$ (at times $t_\alpha,t_\beta,\cdots,t_\epsilon$) with respective outcomes $r_\alpha,r_\beta,\cdots,r_\epsilon$ and similarly $B=(\zeta, r_\zeta; \eta, r_\eta; \ldots;\nu, r_\nu)$; the quantity $\Pi(B|A)$ represents the probability of realization of $B$, given $A$. (We have changed HVW's notation $\Pi(A|B)$ to  $\Pi(B|A)$ to bring it in correspondence with standard usage in probability theory.) The ordering of times ($t_\alpha,t_\beta,\cdots,t_\epsilon,t_\zeta,t_\eta,\ldots,t_\nu$) is arbitrary. The objects $\Pi(B|A)$ are quite general and can be employed in classical as well as quantum mechanics. All measurements refer to external observers. 

An invariance transformation is defined as an invertible mapping 
\begin{equation}
\label{3.1}			
\alpha \leftrightarrow \overline{\alpha},\beta\leftrightarrow\overline{\beta},\cdots
\end{equation}				       
(recall that the symbol $\alpha$ implicitly includes the time $t_\alpha$ of the measurement $\alpha$) which leaves the $\Pi$ function invariant.
\begin{equation}
\label{3.2}	
\Pi(\overline{\zeta},r_{\overline{\zeta}};\cdots;\overline{\nu},r_{\overline{\nu}})=\Pi \left(\zeta,r_\zeta;\ldots;\nu,r_\nu\right).
\end{equation}				 	
Geometric invariances are the subclass of these mappings which correspond to transformations between reference frames.

HVW explored some consequences of this definition in the classical (Newtonian) mechanics of point particles, relativistic mechanics of point particles and in quantum theory. For the expressions for the $\Pi$-functions in these cases we refer to HVW [20]. From our point of view, the main result is the generalization of Wigner's theorem given below. In the statement of this theorem, the symbol $\alpha$ of eq.(\ref{3.1}) is taken to represent a decision measurement (Yes-No Experiment) represented by a pair $(P_\alpha,t_\alpha)$ where $P_\alpha$ is a $1$-dimensional projection operator and $t_\alpha$ is the time of measurement.
		
Generalized Wigner's Theorem: A mapping of decision measurements onto decision measurements $(P_\alpha,t_\alpha \rightarrow \overline{P}_\alpha,\overline{t}_\alpha)$ will leave the $\Pi$ function invariant if and only if the following two conditions apply:
\begin{enumerate}
\item[(a)]$\overline{P}_\alpha = U P_\alpha U^{-1}$ where $U$ is a unitary or antiunitary operator mapping bijectively coherent subspaces (of the quantum mechanical Hilbert space $\mathcal{H}$ of the system in question) onto coherent subspaces.
\item[(b)]The time order of measurements is either preserved or reversed by the mapping.
\end{enumerate}

We draw two conclusions from the foregoing:
\begin{enumerate}
\item[(i)]Invariance of diagonal elements $d(\alpha,\alpha)$ (see eqs.(\ref{2.12}),(\ref{3.2})) of decoherence functionals must be a part of our definition of symmetry (or an implication of it).
\item[(ii)]One should generally expect a two-way connection between Wigner type symmetries and symmetries defined in the language of histories.
\end{enumerate}

It is of some relevance here to note the distinction between symmetries of the  formalism (unitary/antiunitary transformations in quantum mechanics, canonical transformations in classical mechanics) and symmetries of dynamics (subclass of symmetries of the formalism leaving the Hamiltonian invariant). Not every symmetry of the formalism need be a symmetry of some given Hamiltonian. For example, the parity operator $P$ given by $(Pf)(x) = f(-x)$ is a unitary operator in $L^{2}(R)$; however, $PHP^{-1}\neq H$ for $H=-\frac{\partial^2}{\partial x^2}+x$.

An interesting point to note in the above theorem is that the invariance condition does \emph{not} imply a symmetry of dynamics contrary to what one might intuitively expect. (After all, histories or the $\Pi$ functions are supposed to contain all information about dynamics.)

This can also be seen explicitly by having a closer look at eq.(\ref{1.2}) for $d(\alpha,\alpha)$. A common unitary transformation on all the ingredients - $\rho(t_0)$, the projectors $\alpha_{t_j}$ and the evolution operators $U(t_j,t_k)$ - leaves $d(\alpha,\alpha)$ invariant (in fact it leaves $d(\alpha,\beta)$ of eq.(\ref{1.5}) invariant); it does not have to leave  the Hamiltonian invariant to achieve this.

Now we take up the treatment of symmetries in the formalism of section \ref{two}.


\subsection{Symmetries in the augmented temporal logic formalism}	

We shall use the following notations:
\begin{enumerate}
\item[] {\sf S} = $(\mathcal{U},\mathcal{T},\sigma)$ {\bfseries\emph{(History System)}}
\item[] $\tilde{\mbox{\sf S}}$ = $(\tilde{\mathcal{U}},\tilde{\mathcal{T}},\tilde{\sigma})$ {\bfseries \emph{(Standardized History System)}}
\item[] ({\sf S}) = $(\mathcal{U},\mathcal{T},\sigma,\Omega,\mathcal{D})$ {\bfseries \emph{(Augmented History System)}}
\end{enumerate}

A {\bfseries \emph{morphism}} from {\sf S} into {\sf S'} is a pair $\Phi=(\Phi_1,\Phi_2)$ of mappings such that
\begin{enumerate}
\item[(i)] $\Phi_1 : \mathcal{T}\rightarrow\mathcal{T}'$ is a psg homomorphism or anti-homomorphism, \emph{i.e.} it satisfies either (a) or (b) below.
\begin{enumerate}
\alphaeqn
\item[(a)] $\tau_1\lhd\tau_2$ implies $\Phi_1(\tau_1)\lhd\Phi_1(\tau_2)$ and 
\begin{equation}
\label{3.3a}		
\Phi_1(\tau_1\circ\tau_2) = \Phi_1(\tau_1)\circ\Phi_1(\tau_2).
\end{equation}				
\item[(b)] $\tau_1\lhd\tau_2$ implies $\Phi_1(\tau_2)\lhd\Phi_1(\tau_1)$ and
\begin{equation}
\label{3.3b}			
\Phi_1(\tau_1\circ\tau_2) = \Phi_1(\tau_2)\circ\Phi_1(\tau_1).
\end{equation}				
\reseteqn
\end{enumerate}
\item[(ii)] $\Phi_2 : \mathcal{U}\rightarrow\mathcal{U}'$ is a psg homomorphism or anti-homomorphism in accordance with (i) (\emph{i.e.} $\Phi_1$ and $\Phi_2$ are either both  homomorphisms or both antihomomorphisms).
\item[(iii)] The following diagram is commutative. 


\begin{equation}\label{3.4}
\begin{CD}
\mathcal{U}@>\Phi_2>>\mathcal{U}' \\
@V{\sigma}VV   @VV{\sigma' \hspace{.4cm} i.e. \hspace{.5cm} \Phi_1\circ
\sigma = \sigma'\circ\Phi_2.}V \\
\mathcal{T}@>>\Phi_1>\mathcal{T}'
\end{CD}
\end{equation}

 
Writing $\Phi_1(\tau)=\tau'$ and $\Phi_2 |\mathcal{U}_\tau =\Phi_{2\tau}$, the restriction of $\Phi_2$ to the space $\mathcal{U}_\tau$, eq.(\ref{3.4}) implies that $\Phi_{2\tau}$ maps $\mathcal{U}_\tau$ into $\mathcal{U}'_{\tau '}$.
\item[(iv)] Each mapping $\Phi_{2\tau} : \mathcal{U}_\tau\rightarrow\mathcal{U}'_{\tau '}$ is a morphism of logics [10] (\emph{i.e.} it is injective, preserves partial order, meet, join, orthocomplementation, and maps null and unit elements onto null and unit elements respectively).
\end{enumerate}

A morphism is called an {\bfseries\emph{isomorphism}} if the mappings $\Phi_1$ and $\Phi_2$ are bijective. An isomorphism of {\sf S} onto itself is called an {\bfseries\emph{automorphism}} of {\sf S}. The family of all automorphisms of {\sf S} forms a group called Aut({\sf S}).

The next step is to obtain, from $\Phi$, an induced morphism $\tilde\Phi$=$(\tilde\Phi_1,\tilde\Phi_2)$ of $\tilde{\mbox{\sf S}}$ onto $\tilde{\mbox{\sf S}}'$. This appears to go through smoothly only if $\Phi$ is an isomorphism which we will henceforth assume it to be.

Since each $\Phi_{2\tau}$ is a bijection preserving the logic structure, in particular the null and unit elements, it is clear that $\Phi_2$ maps null history filters onto null history filters (and \emph{vice versa}) and unit history filters onto unit history filters (and \emph{vice versa}). It follows that $\Phi_2$ induces a bijective mapping $\tilde{\Phi}_2$ of $\tilde{\mbox{\sf S}}$ onto $\tilde{\mbox{\sf S}}'$ which maps $\tilde{0}$ to $\tilde{0}'$, $\tilde{1}$ to $\tilde{1}'$ and typical elements to typical elements (and \emph{vice versa}); in fact, it is a psg isomorphism.

It should also be clear that $\Phi_1$ induces a psg isomorphism $\tilde{\Phi}_1$ of $\tilde{\mathcal{T}}$ onto $\tilde{\mathcal{T}}'$ and that the pair $\tilde{\Phi} = (\tilde{\Phi}_1, \tilde{\Phi}_2)$ is an isomorphism of $\tilde{\mbox{\sf S}}$ onto $\tilde{\mbox{\sf S}}'$.

The isomorphism $\tilde{\Phi}_2$ preserves the operations $\leq$, $\bot$, $\oplus$ and $\wedge$ defined on $\tilde{\mathcal{U}}$. The condition of weak disjointness is also preserved. It is now not difficult to see that we have an induced mapping $\underline{\Phi}_2:\Omega\rightarrow\Omega'$ mapping $\underline{0}$ to $\underline{0}'$, $\underline{1}$ to  $\underline{1}'$ and generic elements to generic elements (and \emph{vice versa}).

All the structural properties going into the definition of various operations in $\Omega$ $(\leq,\bot,\oplus)$ are preserved by $\underline{\Phi}_2$. In particular 	
\begin{enumerate}
\item[(a)] $\underline\alpha\bot\underline\beta$\quad if and only if \quad $\underline{\Phi}_2(\underline\alpha)\bot\underline{\Phi}_2(\underline\beta)$
\item[(b)] $\underline{\Phi}_2(\underline\alpha\oplus\underline\beta)=\underline{\Phi}_2(\underline\alpha)\oplus \underline{\Phi}_2(\underline\beta)$
\item[(c)] $\underline{\Phi}_2(\neg\underline\alpha) = \neg \underline{\Phi}_2(\underline\alpha)$.
\end{enumerate}
Henceforth we restrict ourselves to the case $\mathcal{U}'=\mathcal{U}$, $\mathcal{T}'=\mathcal{T}$; this implies $\tilde{\mathcal{U}}'=\tilde{\mathcal{U}}$, $\tilde{\Omega}'=\tilde{\Omega}$ \emph{etc}.

To obtain the transformation law of decoherence functionals (when an explicit expression for d($\underline\alpha,\underline\beta$) in terms of initial state, evolution maps and $\underline\alpha$, $\underline\beta$ is given), we need transformation laws of states and evolution maps.

A state $p_\tau\in\mathcal{S}(\tau)$ transforms under $\Phi$ to a state $p'_{\tau '}\in\mathcal{S}(\tau ')$ (where $\tau '=\Phi_1(\tau)$) is given by
\begin{equation}
\label{3.5}		
p'_{\tau '}(\beta)=p_\tau\left[\Phi^{-1}_{2\tau}(\beta)\right]\quad\mbox{for all}\,\beta\in\mathcal{U}'_{\tau '}.				
\end{equation}				
It is easily seen from eq.(\ref{3.5}) that the mapping $p_\tau\rightarrow p'_{\tau '}$ preserves convex combinations; in particular, it transforms pure states to pure states (and  \emph{vice versa}).

An observable $A_\tau\in\mathcal{O}(\tau)$ transforms under $\Phi$ to $A'_{\tau '}\in\mathcal{O}(\tau ')$ given by
\begin{equation}
\label{3.6}		
A'_{\tau '}(E) = \Phi_{2\tau}\left[A_\tau(E)\right]\quad \mbox{for all}\, 
E\in B(R).
\end{equation}				

Given $\tau_1\lhd\tau_2$ in $\mathcal{N}(\mathcal{T})$ and an evolution map $V(\tau_2, \tau_1): \mathcal{U}_{\tau_1}\rightarrow\mathcal{U}_{\tau_2}$, $\Phi$ induces an evolution map $V'(\tau'_2 ,\tau_1'): \mathcal{U}'_{\tau'_1}\rightarrow\mathcal{U}'_{\tau_2 '}$ such that the following diagram is commutative:


\begin{displaymath}			
\begin{CD}
\mathcal{U}_{\tau_1}@>V(\tau_2,\tau_1)>>\mathcal{U}_{\tau_2} \\
@V{\Phi_{2\tau_1}}VV   @VV{\Phi_{2\tau_2}} V \\
\mathcal{U}_{\tau'_1}@>>V'(\tau'_2,\tau'_1)>\mathcal{U}_{\tau'_2}
\end{CD}
\end{displaymath}			
\begin{eqnarray}
\label{3.7}		
\mbox{\it i.e.} & V'(\tau'_2,\tau'_1)\circ\Phi_{2\tau_1} = \Phi_{2\tau_2}
\circ V(\tau_2,\tau_1) &	
\end{eqnarray}				


Given an expression for $d=d_{p_0,V}:\Omega\times\Omega\rightarrow C$, the pair $\Phi$=($\Phi_1$, $\Phi_2$) induces a map $\Phi_3 : \mathcal{D}\rightarrow\mathcal{D}$, given by
\begin{equation}
\label{3.8}		
\Phi_3 \left[d_{p_0,V}\right] = d_{p'_{0'},V'}
\end{equation}				
where $	p'_{0'}$, $V'$ are given by (\ref{3.5}) and (\ref{3.7}). If such an expression is not given, $\Phi_3$ may formally  be treated as an independent mapping for the purpose of the definition of symmetry given below.

A {\bfseries\emph{symmetry operation}} for the augmented history system ({\sf S}) (see notation above) is a triple $\underline\Phi$ = ($\underline\Phi_1$,$\underline\Phi_2$,$\underline\Phi_3$) such that
\begin{itemize}							
\item[(i)]{ $\Phi$ = ($\Phi_1$,$\Phi_2$) is an automorphism of {\sf S}};
\item[(ii)] {$\Phi_3$ : $\mathcal{D} \rightarrow \mathcal{D}$ satisfies the condition
\begin{equation}
\label{3.9}		
\mbox{Re}\left[\underline\Phi_3(d)\left(\underline\Phi_2(\underline\alpha),\underline\Phi_2(\underline\beta)\right)\right] = \mbox{Re}[d(\underline\alpha,\underline\beta)] \hspace{1cm} \mbox{for all } \underline\alpha,\underline\beta \in \Omega.	     
\end{equation}}				
\end{itemize}   
Note that the condition (\ref{3.9}) implies preservation of the decoherence condition (\ref{2.11}) as well as the probability expression of eq.(\ref{2.12}).  A symmetry operation is called \emph{orthochronous} if the mappings ($\underline\Phi_1$,$\underline\Phi_2$) are homomorphisms and \emph{non-orthochronous} if they are antihomomorphisms. They are distinguished by the fact that the former preserves the `temporal order' of `events' while the latter reverses it.				

A symmetry operation as defined above should be understood to be one in the sense of a symmetry of the formalism. A symmetry of dynamics would be a member of the subclass of these symmetries leaving the evolution map $V(.,.)$ invariant, \emph{i.e.} satisfying the condition (\ref{3.7})
\begin{equation}
\label{3.10}		
V'(\tau_2,\tau_1) = V(\tau_2,\tau_1),
\end{equation}				
for all $\tau_2,\tau_1$ such that $\tau_1\lhd\tau_2$. This equation can be  meaningful only if the two sides define mappings between the same spaces; this implies $\mathcal{U}'_{\tau_2} \approx \mathcal{U}_{\tau_2}$ and  $\mathcal{U}'_{\tau_1} \approx \mathcal{U}_{\tau_1}$ where $\approx$ indicates isomorphism (of logics). Since $\tau_1$ and $\tau_2$ are fairly arbitrary (subject only to the condition $\tau_1\lhd\tau_2$), it appears that, in the context of the class of theories being discussed, symmetries of dynamics are definable only for the subclass in which all $\mathcal{U}_\tau$'s are isomorphic.

Remarks: (1) Given a pair of spaces $(\mathcal{U},\mathcal{T})$, there may be more than one possible candidates for the objects $d_{p,V}$. The family of transformations constituting symmetries will then be correspondingly different for different $d_{p,V}$'s. This is due to the invariance condition (\ref{3.9}) which varies with the choice of $d_{p,V}$.

(2) The choice (\ref{3.9}) for the invariance condition was motivated by the need to preserve the decoherence condition (\ref{2.11}) and the probability expression (\ref{2.12}). Instead of eq.(\ref{2.11}), a stronger condition is often employed [4-6], namely
\begin{equation}
\label{3.11}		
d(\alpha,\beta) = 0 , \hspace{1in} \alpha\neq\beta.
\end{equation}				
The appropriate invariance condition replacing eq.(\ref{3.9}) would then be
\begin{equation}
\label{3.12}		
\underline\Phi_3 (d)\left(\underline{\Phi}_2(\underline{\alpha}),\underline{\Phi}_2(\underline{\beta})\right) = d \left(\underline{\alpha},\underline{\beta}\right).
\end{equation}				
In fact, as we shall see in section \ref{four} and \ref{five}, the symmetries which correspond to the traditional symmetries of nonrelativistic quantum mechanics and those in classical mechanics satisfy the stronger invariance condition (\ref{3.12}).


\subsection{Physical equivalence of histories}				
Gell-Mann and Hartle [14] have emphasized the need to understand the nature of physical equivalence between sets of (coarse-grained) histories of a closed system as a prerequisite for a clear understanding of some fundamental questions like what would it mean for the universe to exhibit essentially inequivalent quasiclassical realms. Assuming, for simplicity, a fixed spacetime geometry permitting foliation in space-like hypersurfaces and that the underlying dynamics of the  universe is governed by a canonical quantum field theory, they have considered, in histories-based version of this dynamics, a few notions of physical equivalence of sets of histories. We shall now show that all these notions reduce to special cases of a simple criterion for physical equivalence which can be stated concisely in terms of the symmetry operations described above.

The obvious guiding principle for such a criterion is that histories related through transformations (of relevant entities) leaving all observable quantities invariant must be treated as physically equivalent. The observable quantities for histories (of a closed system) are the probabilities of (decoherent) histories and (in quasitemporal theories [8]) `temporal order' of `events'. The following criterion for physical equivalence suggest itself: All histories related to each other through orthochronous symmetry operations are physically equivalent.		
				
We now take up various notions of physical equivalence considered by GH [14]. Since, in our description of symmetry in terms of a triple $(\Phi_1,\Phi_2,\Phi_3)$, $\Phi_3$ is determined (given a decoherence functional like (\ref{1.5})) in terms of $\Phi_1$ and $\Phi_2$, it is adequate to give $\Phi_1$ and $\Phi_2$ corresponding to the various GH notions of physical equivalence. (At this point, readers are advised to go through section \ref{four} up to eq.(\ref{4.18}).)
\begin{itemize}
\item[(i)]{A fixed unitary transformation (say $U$) applied to all operators (including the initial density operator). In this case, we have $\Phi_1$ = identity and $\Phi_2$ is the automorphism given by eq.(\ref{4.18}) below (combined with eqs. (\ref{4.10}) and (\ref{4.11})). }
\item[(ii)]{ The same operator described in terms of fields at different times (using Heisenberg equations of motion), say at times $t$ and $t+a$. In this case, we have
\begin{eqnarray}
\label{3.13}		
& \Phi_1(t) = t+a; & \Phi_{2t} = V(t+a,t); 		
\end{eqnarray}				
where $V(t,t')$ is the evolution operator in our formalism (given by the $\tilde U$ of eqs. (\ref{4.18}),(\ref{4.10}) and (\ref{4.11}) corresponding to $U = U(t,t') = \exp [-iH(t-t')/\hbar]$).}
\item[(iii)]{Histories related through field redefinitions: Given a field transformation $(\phi,\pi)\rightarrow(\phi',\pi')$, the two histories involve, typically, observables $A$ and $A'$ related through equations like $A[\phi,\pi]=A'[\phi',\pi']$. Since, as operators, $A$ and $A'$ are identical (and have, therefore, identical spectral projectors), the mappings $\Phi_1$ and $\Phi_2$ are both identity maps. As an example, let $\phi'=U^{-1}\phi U, \pi'=U^{-1}\pi U, A=\int\pi\dot{\phi}$; then 
\begin{equation}
\label{3.14}		
A[\phi,\pi]\equiv\int\pi\dot{\phi}dx = U [\int\pi'\phi'dx]U^{-1} \equiv A'[\phi',\pi'].
\end{equation}}				
\item[(iv)]{ Another notion of physical equivalence described in section II-F of GH [14] states that two histories described, respectively, by triples $(\{C_\alpha\},H,\rho)$ and $(\{\tilde{C}_\alpha\},\tilde{H},\tilde{\rho})$ (where $C_\alpha$ are the class operators of the type of eq.(\ref{1.3}), $H$ is the Hamiltonian and $\rho$ is the density operator representing the initial state) are physically equivalent if there exist canonical pairs ($\phi$,$\pi$) and ($\tilde\phi$,$\tilde\pi$) [each satisfying the standard canonical commutation relations (CCR)] such that the relevant operators in one history have the same expressions in terms of ($\phi$,$\pi$) as those in the other have in terms of ($\tilde\phi$,$\tilde\pi$). Assuming as above, that the theory in question has a concrete expression for the decoherence functional $d(\alpha,\beta)$ and denoting the histories corresponding to the two triples above as $\alpha$ and $\tilde\alpha$ respectively, we must have 
\begin{equation}
\label{3.15}		
d(\tilde\alpha,\tilde\beta)=d(\alpha,\beta).
\end{equation}				
This is because, when the two sides are expressed in terms of the canonical pairs $(\phi,\pi)$ and $(\tilde\phi,\tilde\pi)$, there is nothing to mathematically distinguish the two sides (apart from some trivial relabelling). In this case, we have $\Phi_1$ = identity and $\Phi_2$ (or $\underline\Phi_2$) precisely the mapping given by $\alpha\rightarrow\tilde\alpha$.}
\end{itemize}
Remark: The criterion of physical equivalence of histories stated above is applicable to closed systems only and is not applicable, for example, to the history version of traditional quantum mechanics in which eq.(\ref{1.2}) is the probability for the history (\ref{1.1}) in which events refer to measurements by an external observer. GH [14] have emphasized that the criterion for physical equivalence of histories have to be different for closed systems (ideally, the universe where an observer or measurement apparatus is a part of the system) and for the `approximate quantum mechanics of a measured subsystem' of the universe. The latter categories of theories (of which standard quantum mechanics is an example) have external observers which employ reference frames; two histories related through a nontrivial transformation within a reference frame (for example, a time translation, a space translation or a spatial rotation) are physically distinguishable.


\subsection{Symmetries in general quasitemporal theories and in Isham-Linden type theories}
In going from the augmented temporal logic formalism to general quasitemporal theories [8] and from there to the Isham-Linden type theories [11], one has to drop some structures along the way. In the first transition, we have to drop the logic structure of $\mathcal{U}_\tau$'s and, in the second, the quasitemporal structure itself. The definition of symmetry in these theories must ensure preservation of the remaining mathematical structure.

A concrete quasitemporal theory must define an embedding of $\mathcal{U}$ in $\Omega$ in concrete terms. Such a theory has associated with it what we have called above an augmented history system, ({\sf S})= $(\mathcal{U},\mathcal{T},\sigma,\Omega,\mathcal{D})$. Here, as noted in the previous section,the triple $(\mathcal{U},\mathcal{T},\sigma)$ is the analogue of $(\tilde\mathcal{U},\tilde\mathcal{T},\tilde\sigma)$ above. A symmetry operation in such a theory may be defined as a triple $\Phi=(\Phi_1,\Phi_2,\Phi_3)$ of (invertible) mappings satisfying the conditions stated above except that
\begin{itemize}								
\item[(i)]the condition (iv) on $\Phi_{2\tau}$ is not relevant and must be dropped.
\item[(ii)] the mapping $\Phi_3$ now cannot be specified as in eq.(\ref{3.8}) and must be kept general.
\end{itemize}

In the Isham-Linden type theories, a symmetry may be defined as a pair $(\underline\Phi_2,\Phi_3)$ of invertible mappings ($\underline\Phi_2:\Omega\rightarrow\Omega$ and $\Phi_3:\mathcal{D}\rightarrow\mathcal{D}$) having the properties as described earlier (recall, in particular, that $\underline\Phi_2$ preserves the mathematical operations in $\Omega$) and satisfying the invariance condition (\ref{3.9}). In each of these theories, the invariance condition (\ref{3.9}), if appropriate, may be replaced by the stronger condition (\ref{3.12}).

If, in any of the theories discussed above, a concrete expression for the decoherence functional is available, the mappings involved in the definition of symmetry can be generally be shown to belong to well defined classes. This was the situation in the work of HVW described above (where we had the $\Pi$ functions instead of the decoherence functional) and prevails in sections \ref{four} and \ref{five} and in the work of Schreckenberg [15] briefly described below.

Isham, Linden and Schreckenberg [16] proved, for the special case of Isham-Linden type theories in which $\Omega$ is the lattice $\mathcal{P}(\mathcal{V})$ of projection operators in a finite dimensional Hilbert space $\mathcal{V}$ (of dimension $>$ 2), that every decoherence functional $d(\underline\alpha,\underline\beta)$ where $\underline\alpha,\underline\beta \in \mathcal{P}(\mathcal{V})$ can be written as
\begin{equation}
\label{3.16}		
d(\underline\alpha,\underline\beta) = \mbox{tr}_{\mathcal{V}\otimes\mathcal{V}}\left[(\underline\alpha\otimes\underline\beta)X\right]
\end{equation}				
where $X$ belongs to a class $\mathcal{X}_\mathcal{D}$ of operators in the space $\mathcal{V}\otimes\mathcal{V}$ satisfying a definite set of conditions of hermiticity, positivity and normalization (chosen so as to have the $d$ of eq.(\ref{3.16}) satisfy the usual conditions). This is the history analogue of the famous Gleason's theorem [10] in standard quantum theory.

In Schreckenberg's paper [15] `physical symmetries' were defined in the framework of theories of the above sort as affine one-to-one maps from $[\mathcal{P}(\mathcal{V})\otimes\mathcal{P}(\mathcal{V})]\times\mathcal{X}_\mathcal{D}$ into itself preserving the quantity on the right hand side of eq.(\ref{3.16}). A Wigner type theorem [21,10] was proved there showing that the `physical symmetries' are in one-to-one correspondence with the so-called `homogeneous symmetries' (those implemented by the operators of the form $U\otimes U$ on $\mathcal{V}\otimes\mathcal{V}$ where $U$ is a unitary or antiunitary operator on $\mathcal{V}$). Here a symmetry operation can be easily seen to be described as a pair ($\underline\Phi_2$,$\Phi_3$) where the mappings $\underline\Phi_2:\mathcal{P}(\mathcal{V})\rightarrow\mathcal{P}(\mathcal{V})$ and $\Phi_3:\mathcal{X}_\mathcal{D}\rightarrow\mathcal{X}_\mathcal{D}$ are those in eqs. (II.17),(II.18) and (III.7) of [15]. In [22], symmetries of individual decoherence functionals (maps $\underline\alpha\rightarrow\underline\alpha'$ = $U\underline\alpha U^{\dagger}$, $\underline\beta\rightarrow\underline\beta'$ satisfying the condition $d(\underline\alpha',\underline\beta')=d(\underline\alpha,\underline\beta)$ for all $\underline\alpha, \underline\beta$ in $\Omega = \mathcal{P}(\mathcal{V})$ for a given $d$) were considered in some detail.


\section{Traditional vs temporal logic descriptions of symmetries in nonrelativistic quantum mechanics}
\label{four}

In this section we shall establish a reciprocal relationship between the traditional description of symmetry operations in nonrelativistic quantum mechanics and those in the present formalism.		
			
Transition from traditional Wigner symmetries to those in the formalism of the previous section is described most transparently in the HPO (History Projection Operator) formalism [8].


\subsection{The HPO formalism}
If, following Isham's suggestion as mentioned in section \ref{one}, histories are to be represented as (multitime) propositions, it is natural to look for a representation of histories as projection operators in some Hilbert space. This was achieved for traditional quantum theory in [8]. In this case, one first constructs the Cartesian product
\begin{equation}
\label{4.1}		
\mathcal{V} = \Pi_{t\in R}\mathcal{H}_t
\end{equation}				
of (naturally isomorphic) copies $\mathcal{H}_t$ of the quantum mechanical Hilbert space $\mathcal{H}$ of the system. Let $w=(w_t)$ be a fixed vector in $\mathcal{V}$ such that $\parallel w_t \parallel=1$ for all $t\in R$. Let $\mathcal{F}$ be the subspace of $\mathcal{V}$ consisting of vectors $v$ such that $v_t = w_t$ for all but a finite set of $t-$values. The scalar product $(.,.)$ on $\mathcal{H}$ induces the following scalar product on $\mathcal{F}$:
\begin{equation}
\label{4.2}		
(v',v)_\mathcal{F} = \Pi_{t\in R} (v'_t,v_t)_{\mathcal{H}_t}.
\end{equation}				
The completion $\tilde{\mathcal{H}}$ of this inner product space is the desired Hilbert space; it is the infinite tensor product
\begin{equation}
\label{4.3}		
\tilde{\mathcal{H}} = \otimes^{w}_{t\in R} \mathcal{H}_t.
\end{equation}				
The history $\alpha$ of eq.(\ref{1.1}) is represented in $\tilde{\mathcal{H}}$ by the homogeneous projection operator 
\begin{equation}
\label{4.4}		
\overline{\alpha}=\alpha_{t_1}\otimes\alpha_{t_2}\otimes\cdots\otimes\alpha_{t_n}.
\end{equation} 				
Given another projection operator $\overline{\beta}$ in $\tilde{\mathcal{H}}$ representing some history $\beta$ such that $\overline{\alpha}$ and  $\overline{\beta}$ are mutually orthogonal, the inhomogeneous projection operator $\overline{\alpha}+\overline{\beta}$ can be taken to represent the history proposition `$\alpha$ or $\beta$'. Histories are referred to as `homogeneous' or `inhomogeneous' depending on whether they are represented by homogeneous or inhomogeneous projection operators. Inclusion of inhomogeneous histories facilitates the introduction of the logical operation of negation of a history proposition. Given $\alpha$ and $\overline{\alpha}$ as above, the projection operator representing the negation of the history proposition $\alpha$ is $\tilde{I}-\overline{\alpha}$ where $\tilde{I}$ is the identity operator on $\tilde\mathcal{H}$.  


\subsection{Symmetries in traditional quantum mechanics in the HPO formalism}

We denote by $\mathcal{P}_1(\mathcal{H})$ the space of one-dimensional projection operators (\emph{i.e.} objects of the form $P_\Psi = |\Psi\rangle\langle\Psi|$ on the quantum mechanical Hilbert space $\mathcal{H}$). In terms of these objects the transition probability formula reads 
\begin{equation}
\label{4.5}
P(|\Psi\rangle\rightarrow|\Phi\rangle)=|\langle\Phi|\Psi\rangle|^2=\mbox{Tr}(P_\Psi P_\Phi).
\end{equation}					
	
According to Wigner's theorem [22,10], given a bijection $\mathcal{P}_1(\mathcal{H}) \rightarrow \mathcal{P}_1(\mathcal{H})$ ($P\rightarrow P'$) such that		
\begin{equation}
\label{4.6}		
\mbox{Tr}(P'_1 P'_2) = \mbox{Tr}(P_1 P_2),
\end{equation}				
there exists a unitary or antiunitary operator $U$ on $\mathcal{H}$ such that 
\begin{equation}
\label{4.7}		
 P'= UPU^{-1}\hspace{2cm}\mbox{for all}\hspace{.2cm} P \in \mathcal{P}_1(\mathcal{H}).
\end{equation}				
Recall that, the action of such a $U$ on $\mathcal{B}(\mathcal{H})$ (the algebra of bounded operators on $\mathcal{H}$) is given by 
\alphaeqn
\begin{eqnarray}
\label{4.8}		
A \rightarrow A' =&  U A U^{-1}  &\mbox{for unitary}\quad U \\
A \rightarrow A' =&  U A^{*} U^{-1} &\mbox{for antiunitary}\quad U
\end{eqnarray}				
\reseteqn
If $A$ is self-adjoint, the expressions on the right side in eqs.(\ref{4.8}-b) are the same; in particular, this is the case for a (general, \emph{i.e.} not necessarily one dimensional) projection operator and for a density operator.

The unitary/antiunitary operator $U$ on $\mathcal{H}$ defines a unitary/antiunitary operator $\tilde{U}$ on $\tilde{\mathcal{H}}$ such that 
\begin{equation}
\label{4.9}
\tilde{U}\left[\Psi_{t_1}\otimes\Psi_{t_2}\otimes\cdots\otimes\Psi_{t_n}\right] 
= U\Psi_{t_1}\otimes U\Psi_{t_2}\otimes\cdots\otimes U \Psi_{t_n}.
\end{equation}				
The homogeneous projection operator $\overline{\alpha}$ of eq.(\ref{4.4}) transforms into
\begin{equation}
\label{4.10}		
\overline{\alpha}' = \tilde{U}\overline{\alpha}\tilde{U}^{-1}=\alpha'_{t_1}\otimes\cdots\otimes\alpha'_{t_n}
\end{equation}				
where
\begin{equation}
\label{4.11}	 	
\alpha'_{t_j} = U \alpha_{t_j} U^{-1}.
\end{equation}				
This action trivially extends to inhomogeneous projectors:
\begin{equation}
\label{4.12}		
\overline{\alpha}+\overline{\beta}\rightarrow\overline{\alpha}'+\overline{\beta}' = \tilde{U}(\overline{\alpha}+\overline{\beta})\tilde{U}^{-1}.
\end{equation}				
Given $\overline{\alpha}$ of eq.(\ref{4.4}) and $\overline{\beta}=\beta_{s_1}\otimes\cdots\otimes\beta_{s_m}$ such that $t_n<s_1$, the composite homogeneous history projector 
\begin{displaymath}			
\overline{\alpha}\circ\overline{\beta}=\alpha_{t_1}\otimes\cdots\otimes \alpha_{t_n}\otimes\beta_{s_1}\otimes\cdots\otimes\beta_{s_m},
\end{displaymath}			
being a homogeneous projector, transforms as in eq.(\ref{4.10}) giving
\begin{equation}
\label{4.13}		
\overline{\alpha}\circ\overline{\beta}\rightarrow \tilde{U}(\overline{\alpha}\circ\overline{\beta})\tilde{U}^{-1} = (\tilde{U}\overline{\alpha}\tilde{U}^{-1})\circ(\tilde{U}\overline{\beta}\tilde{U}^{-1}).
\end{equation}				
The transformation law (\ref{4.10}), therefore, defines an automorphism of the psg $\mathcal{K}_2$ of section \ref{one} (which is nothing but the space $\mathcal{U}$ of section \ref{two} in the present case).

In writing eq.(\ref{4.13}), we have implicitly assumed that there is no transformation of time involved. There may, in general, be a transformation of the time variable also involved,
\begin{equation}
\label{4.14}	
t\rightarrow t' = f(t).		
\end{equation}				
If $t'$ is to serve as a time variable, the function $f$ must be monotone. There are two possibilities:
\begin{enumerate}
\item[(i)] $f$ is monotone increasing. In this case $t_i < t_j$ implies $t'_i < t'_j$. Given temporal supports $A = (t_1,\ldots,t_n)$ and $B = (s_1,\ldots,s_m)$ such that $A\lhd B$ (corresponding to $t_n < s_1$) we have
\alphaeqn
\begin{equation}
\label{4.15a}		
A'=(t'_1,\ldots,t'_n) \hspace{1cm} B' = (s'_1,\ldots,s'_m)
\end{equation}				
and
\begin{equation}
\label{4.15b}		
A' \lhd B' , \hspace{2cm}	(A\circ B)' = A'\circ B'
\end{equation}				
\reseteqn
giving an automorphism of the psg $\mathcal{K}_1$ (which is the space $\mathcal{T}$ in the present case).
\item[(ii)] $f$ is monotone decreasing (time reversal $t'=-t$ is an important special case of this). In this case $t_i < t_j$ implies $t'_j < t'_i$ and we have
\alphaeqn
\begin{equation}
\label{4.16a}		
B' \lhd  A'
\end{equation}				
\begin{equation}
\label{4.16b}
(A\circ B)'  =  B' \circ A'
\end{equation}				
\reseteqn
giving an anti-automorphism of $\mathcal{K}_1$. 
\end{enumerate}

In this case eqs.(\ref{4.16a}) implies that eq.(\ref{4.13}) must be replaced by an anti-automorphism of $\mathcal{U} = \mathcal{K}_2$:
\begin{equation}
\label{4.17}		
\overline{\alpha}\circ\overline{\beta}\rightarrow \tilde{U}(\overline{\alpha}\circ\overline{\beta})\tilde{U}^{-1} = (\tilde{U}\overline{\beta}\tilde{U}^{-1})\circ(\tilde{U}\overline{\alpha}\tilde{U}^{-1}).
\end{equation}			
Summarizing, we have shown that the Wigner symmetry implemented as in eq.(\ref{4.7}) along with the transformation (\ref{4.14}) of time implies, for the history system ($\mathcal{U}$, $\mathcal{T}$, $\sigma$) = ($\mathcal{K}_2$, $\mathcal{K}_1$, $\sigma$) (with the homomorphism $\sigma$ given by eq.(\ref{1.13})), the following
\begin{enumerate}
\item[(i)] an automorphism or antiautomorphism $\Phi_1$ of $\mathcal{T}$ (given by $\Phi_1(A) = A'$);
\item[(ii)] an automorphism or antiautomorphism $\Phi_2$ of $\mathcal{U}$ (in accordance with (i)) given by
\begin{equation}
\label{4.18}		
\Phi_2(\overline{\alpha}) = \tilde{U}\overline{\alpha}\tilde{U}^{-1}
\end{equation}				
\item[(iii)] projections (from $\mathcal{U}$ to $\mathcal{T}$) implied by $\sigma$ are preserved by the mappings $\Phi_1$ and $\Phi_2$ making the following diagram commutative:


\begin{equation}
\label{4.18.5}
\begin{CD}
\mathcal{U}@>\Phi_2>>\mathcal{U}' \\
@V{\sigma}VV  @VV{\sigma \hspace{.3cm} i.e. \hspace{.4cm} \Phi_1\circ\sigma
=\sigma\circ\Phi_2.}V \\
\mathcal{T}@>>\Phi_1>\mathcal{T}'	
\end{CD}
\end{equation}				


Eq.(\ref{4.18.5}) implies that, for any $t\in R$ (=$\mathcal{N}(\mathcal{T}$) in the present case), $\Phi_2$ maps $\mathcal{U}_t=\mathcal{P}(\mathcal{H}_t)$ into $\mathcal{U}_{t'} = \mathcal{P}(\mathcal{H}_{t'})$ where $t'=\Phi_1(t)=f(t)$.
\item[(iv)] for each $t\in R = \mathcal{N}(\mathcal{T})$, the mapping $\Phi_{2t}= \Phi_2 \mid\mathcal{U}_t$ is an isomorphism of the logic $\mathcal{U}_t =\mathcal{P}(\mathcal{H}_t)$ onto  $\mathcal{U}_{t'} = \mathcal{P}(\mathcal{H}_{t'})$.
\item[(v)] Eq.(\ref{4.12}) and (iv) above imply that the mapping $\Phi_2$ on $\mathcal{U}=\mathcal{K}_2$ extends to a bijective mapping $\underline{\Phi}_2$ on the space $\Omega$ of inhomogeneous projectors onto itself preserving the lattice operations in it.
\end{enumerate}				
					
Note: Whether $\Phi_1$ and $\Phi_2$ are automorphisms or antiautomorphisms depends only on whether $f(t)$ in eq.(\ref{4.14}) is monotone increasing or decreasing and not on whether $U$ (and corresponding $\tilde{U}$) is unitary or antiunitary. We can very well have a situation where, for example, $U$ is antiunitary, $\Phi_1$ is the identity mapping and $\Phi_2$ is an automorphism.

The transformation law of the decoherence functionals $d(\alpha,\beta)$ of eq.(\ref{1.5}) following from that of the initial state $\rho(t_0)$ and the evolution operator $U(t',t)$ given by eqs.(\ref{4.8}-b) gives $\rho(t_0)\rightarrow U\rho(t_0) U^{-1}$ and
\alphaeqn
\begin{equation}
\label{4.19a}		
C_\alpha \rightarrow 
\begin{cases}
U C_\alpha U^{-1}&	\text{for $U$ unitary}, \\
U C^{\dagger}_\alpha U^{-1}&	\text{for $U$ antiunitary}.
\end{cases}
\end{equation}				
which implies that 
\begin{equation}
\label{4.19b}		
d(\alpha,\beta)\rightarrow
\begin{cases}
d(\alpha,\beta)& 	\text{for $U$ unitary}, \\
d(\alpha,\beta)^{*}&\text{for $U$ antiunitary}.
\end{cases}
\end{equation}				
\reseteqn
Eq.(\ref{4.19b}) describes the mapping $\Phi_3$ of eq.(\ref{3.8}) and is clearly consistent with eq.(\ref{3.9}). 

We have seen above that a traditional Wigner symmetry (supplemented with the transformation (\ref{4.14}) of time) implies, in the history version, a symmetry of the type described in section \ref{three}. We now consider the reverse connection.


\subsection{Recovering Wigner symmetries from those in the temporal logic formalism}

The simplest way to obtain Wigner symmetries from those of section \ref{three} is to note that, since, for each $t\in R = \mathcal{N}(\mathcal{T})$, $\mathcal{U}_t\approx\mathcal{P}(\mathcal{H})$ where $\mathcal{H}$ is the quantum mechanical Hilbert space of the system, a symmetry of section \ref{three} implies an automorphism of $\mathcal{P}(\mathcal{H})$ which, in turn, by theorem (4.27) of [10], implies the existence of a unitary/antiunitary mapping on $\mathcal{H}$.

It is interesting to note that, in the above description of the `reverse transition', the condition (\ref{3.9}) in the definition of symmetry did not play any role. Another interesting point to note is that the argument is independent of the nature of `time'; it goes through if $\mathcal{K}_1$ is replaced by a general space $\mathcal{T}$ of temporal supports consistent with axioms $A_1$ and $A_2$ of section \ref{two}. This fact will be used in section \ref{seven}.

A pedagogically simpler route to recover Wigner symmetries from those of section \ref{three} is to obtain from the latter, the condition of invariance of transition probabilities (eq.(\ref{4.6})) and then appeal to Wigner's theorem. To this end (noting that $\mathcal{U}$, $\mathcal{T}$ and $\Omega$ here are the same as in the previous subsection) we apply eq.(\ref{3.9}) to the $\underline{\beta}=\underline{\alpha}={\alpha}$ where $\alpha=\alpha_{t_1}$ is a single time history (\emph{i.e.} a projection operator on $\mathcal{H}$). From eqs.(\ref{1.3}) and (\ref{1.5}) we have $C_\alpha = \alpha_{t_1} U(t_1,t_0)$ and
\begin{eqnarray}
\label{4.20}		
d(\alpha,\alpha)&=&\mbox{Tr}\left[C_\alpha\rho(t_0)C^{\dagger}_\alpha\right]\nonumber \\
 & = & 	\mbox{Tr}\left[\alpha_{t_1} U(t_1,t_0) \rho(t_0) U(t_1,t_0)^{-1}\right].
\end{eqnarray}				
Writing $\Phi_1(t_j)=t'_j$ ($j=0,1$), and $\Phi_2(\alpha)=\alpha' =\alpha'_{t'_j}$, the condition $d'(\alpha',\alpha')=d(\alpha,\alpha)$ gives
\begin{equation}
\label{4.21}		
\mbox{Tr}\left[\alpha'_{t'_1}U'(t'_1,t'_0)\rho'(t'_0)U'(t'_1,t'_0)^{-1}\right] 
= \mbox{Tr}\left[\alpha_{t_1} U(t_1,t_0) \rho(t_0) U(t_1,t_0)^{-1}\right].		
\end{equation}				
Putting $t_1=t_0$ in eq.(\ref{4.21}), we get				
\begin{equation}
\label{4.22}		
\mbox{Tr}\left[\alpha'_{t'_0}\rho'(t'_0)\right]=\mbox{Tr}\left[\alpha_{t_0}\rho(t_0)\right].
\end{equation}		
Taking $\rho(t_0)$ to be a pure state (one dimensional projection operator), $\rho'(t'_0)$ must also be a pure state. (See the statement after eq.(\ref{3.5}).) Similarly, taking $\alpha_{t_0}$ to be one dimensional projector, $\alpha'_{t'_0}$ must also be a one dimensional projector because the mapping $\Phi_{2t_0}$ preserves the logic structure of $\mathcal{U}_{t_0}=\mathcal{P}(\mathcal{H})$. With these choices, eq.(\ref{4.22}) reduces to an equation of the form (\ref{4.6}). Wigner's theorem now does the rest.


\section{Traditional vs temporal logic descriptions of symmetries in the history version of classical mechanics}
\label{five}

For classical mechanics of a system with phase space $\mit\Gamma$, the constructions in [12] were given for the case $\mathcal{T}=\mathcal{K}_1$ (which implies $\mathcal{N}(\mathcal{T})=R$, the traditional space for the flow of time). For each $t\in R$, the space $\mathcal{U}_t$ is an isomorphic copy of $B(\mit\Gamma)$, the Boolean logic of Borel subsets of $\mit\Gamma$. An element of $\mathcal{U}$ is of the form (considering, for simplicity, history filters with finite temporal supports only)
\alphaeqn
\begin{equation}
\label{5.1a}		
\alpha = \{\alpha_{t_1},\alpha_{t_2},\ldots,\alpha_{t_n}; t_1 < t_2 < \ldots < t_n, \alpha_{t_i}\in\mathcal{U}_{t_i}\}
\end{equation}			
which is represented, in the notation of section \ref{two} as
\begin{equation}
\label{5.1b}		
\alpha = \alpha_{t_1}\circ\alpha_{t_2}\cdots\alpha_{t_n}.
\end{equation}				
\reseteqn
Construction of $\tilde{\mathcal{U}}$ and $\Omega$ and descriptions of operations/relations in them are straightforward.

States at any time $t$ are the probability measures on the measurable space $(\mit\Gamma,B(\mit\Gamma))$. Observables at any time $t$ are maps $A_t:B(R)\rightarrow B(\mit\Gamma)$ with the properties as stated in section \ref{two}. Temporal evolution is given by measurable maps $V(t',t) :\mathcal{U}_t\rightarrow\mathcal{U}_{t'}$ which we assume to be bijective. Since single points of $\mit\Gamma$ are elements of $B(\Gamma)$, this defines a bijective map of $\mit\Gamma$ onto itself which we also denote as $V(t',t)$. Given $t_0<t_1<t_2<\ldots <t_n$ and $\xi_0\in\mit\Gamma$ (considered as the collection of single point subsets of $\mathcal{U}_{t_0}$), let $\xi(t_j;\xi_{t_0})$ be the point of $\mit\Gamma$ (considered as the collection of single point subsets of $\mathcal{U}_{t_j}$) given by
\begin{equation}
\label{5.2}		
\xi(t_j;\xi_{t_0}) = \left[ V(t_j,t_{j-1})\cdot V(t_{j-1},t_{j-2})\cdot\cdots\cdot V(t_2,t_1)\cdot V(t_1,t_0)\right]\left(\xi_{t_0}\right).		
\end{equation}
Given a history $\alpha$ as above and another history $\beta=\beta_{t_1}\circ\beta_{t_2}\circ\cdots\circ\beta_{t_m}$, we define
\begin{equation}\label{5.3}
d(\alpha,\beta) = p_{t_0}(E_{\alpha\beta}) =\int_{\mit\Gamma} dp_{t_0}(\xi_{t_0}) K^{\alpha\beta}(\xi_{t_0}),
\end{equation}
where $E_{\alpha\beta}$ is the subset of $\mit\Gamma$ consisting of those points $\xi_{t_0}\in\mit\Gamma$ for which $\xi(t_j;\xi_{t_0})$ lies in $\alpha_{t_j}\cap\beta_{t_j}$, for each $j = 1,2,\ldots ,n$ and $K^{\alpha\beta}$ is the characteristic function of $E_{\alpha\beta}$. Eq.(\ref{5.3}) serves  to define the decoherence functional $d$($=d_{p_0,V}$) for any pair of finite histories. (The general case of two finite histories with different temporal supports can be reduced to the simpler case of common temporal supports by taking some of the $\alpha_{t_i}$ and/or $\beta_{t_j}$ equal to $\mit\Gamma$.)	

In the traditional formalism of Hamiltonian dynamics (in the setting of symplectic manifolds), symmetries of the formalism are the canonical transformations (diffeomorphisms of $\Gamma$ preserving the symplectic form). The histories-based formalism, however, has nothing to do with smooth structures and infinitesimal versions of dynamics (Hamilton's equations). It operates in the more general framework of topological spaces and employs objects like Borel sets and Borel measurable evolution maps. The symmetries of the formalism in the present context, therefore, are bijective Borel measurable maps of $\mit\Gamma$ onto itself. 

After these preliminaries, we now consider the two-way connection between symmetries as mentioned above.
\begin{enumerate}
\item[(1)] Given an invertible transformation of the time variable (eq.(\ref{4.14})) and a bijective Borel measurable mapping $F:\mit\Gamma\rightarrow\mit\Gamma$ $(\xi\rightarrow\xi' = F(\xi))$ we construct a symmetry operation $\underline{\Phi}=(\Phi_1,\Phi_2,\Phi_3)$ as follows:
\begin{enumerate}
\item[(i)] An automorphism/antiautomorphism $\Phi_1$ of $\mathcal{K}_1$ is constructed as in section IV.
\item[(ii)] The mapping $\Phi_{2t} : B(\mit\Gamma)\rightarrow B(\mit\Gamma)$ defined by
\begin{equation}
\label{5.4}		
\Phi_{2t}(A) = F(A) 	\hspace{2cm} \mbox{for all } A\in B(\mit\Gamma)
\end{equation}				
is an automorphism of the logic $B(\mit\Gamma)$ [10]. Note that the mapping $\Phi_{2t}$ is the same for all $t$.
\item[(iii)] Given $\alpha=(\alpha_{t_1},\alpha_{t_2},\ldots,\alpha_{t_n})\in\mathcal{U}$, we have
\begin{equation}
\label{5.5}		
\Phi_2(\alpha) =\left(\Phi_{2t_1}(\alpha_{t_1}),\ldots,\Phi_{2t_n}(\alpha_{t_n})\right).
\end{equation}
\item[(iv)] The verification that the pair $\Phi=(\Phi_1,\Phi_2)$ is an automorphism of the history system {\sf S} = $(\mathcal{U},\mathcal{T},\sigma)$ is straightforward.
\item[(v)] The transformation laws of the objects $p_{t_0}$ and $V_{t,s}=V(t,s)$ appearing in the classical decoherence functional eq.(\ref{5.3}) are obtained using eqs. (\ref{3.4}), (\ref{3.5}) and (\ref{5.4}). This gives, for the former,
\begin{equation}
\label{5.6}
p'_{t'_0}(\beta)=p_{t_0}\left[\Phi_{2t_0}^{-1}(\beta)\right]=p_{t_0}\left[F^{-1}(\beta)\right]
\end{equation}				
and, for the latter,
\begin{equation}
\label{5.7}		
V'_{t'_2,t'_1}(\xi) = F\left[V_{t_2,t_1}\left(F^{-1}(\xi)\right)\right].
\end{equation}				
The transformed decoherence functional is given by 
\begin{equation}
\label{5.8}		
d'(\alpha',\beta')=p'_{t_0}(E'_{\alpha'\beta'})=p_{t_0}[ F^{-1}(E'_{\alpha'\beta'})].
\end{equation}
Now, since the mappings involved are invertible, we have
\begin{eqnarray}
\label{5.9}
E'_{\alpha'\beta'}&=&\{\xi'_0 \in\mit\Gamma;\xi'(t'_j;\xi'_0)\in\alpha'_{t'_j}\cap\beta'_{t'_j}, j=1,\ldots,n\} \nonumber \\
& = &\{\xi_0 \in\mit\Gamma;\xi(t_j;\xi_0)\in\alpha_{t_j}\cap\beta_{t_j}, j=1,\ldots,n\} \nonumber \\
& = & F(E_{\alpha\beta}).
\end{eqnarray}				
Eq.(\ref{5.8}) now gives
\begin{equation}
\label{5.10} 		
d'(\alpha',\beta') = p_{t_0}(E_{\alpha\beta}) = d(\alpha,\beta).
\end{equation}
which verifies eq.(\ref{3.9}) in the present case.
\end{enumerate}
\item[(2)]Given a symmetry operation $\underline{\Phi}=(\Phi_1,\Phi_2,\Phi_3)$ we recover the maps $f$ and $F$ as follows:
\begin{enumerate}
\item[(i)]The restriction of $\Phi_1$ to $\mathcal{N}(\mathcal{T})=R$ fixes the map $f$.
\item[(ii)]$\Phi_2$ determines, for each $t\in\mathcal{N}(\mathcal{T})=R$, $\Phi_{2t}:B(\mit\Gamma)\rightarrow B(\mit\Gamma)$ which, when restricted to single point sets, determines a mapping $F:\mit\Gamma\rightarrow\mit\Gamma$ which is bijective and Borel measurable.
\end{enumerate}
\end{enumerate}


\section{Conservation laws}
\label{six}

We next consider conservation laws in the present formalism. Since we have the concept of evolution defined in the formalism, it is natural to define conservation of an observable in terms of its preservation under evolution. To do this, however, we shall need to compare elements of $\mathcal{U}_\tau$'s for different $\tau$'s. It follows that a primary requirement for the definition of conservation laws is that $\mathcal{U}_\tau$'s for different $\tau$'s be isomorphic. We shall henceforth assume this in this section and identify $\mathcal{U}_\tau$'s for all $\tau\in\mathcal{N}(\mathcal{T})$.

Given $\tau\lhd\tau'$, we say an observable $A\in\mathcal{O}(\tau)$ is conserved under the evolution $V(\tau',\tau) : \mathcal{U}_\tau\rightarrow\mathcal{U}_{\tau'}$ if
\begin{equation}
\label{6.1}		
V(\tau',\tau)(A(E)) = A(E) \hspace{2cm} \mbox{for all } E\in B(R).
\end{equation}				
We shall now formulate an alternative definition of conservation law in terms of equality of probabilities of appropriate single `time' histories. Let $\tau_0\lhd\tau\lhd\tau'$ and suppose a prescription is given to construct a decoherence functional $d_{p_0,V}$ in terms of an initial state $p_0$ at `time' $\tau_0$ and the evolution map $V(.,.)$. Let $\alpha$ and $\beta$ be the single `time' histories given by
\begin{eqnarray}
\label{6.2}		
\alpha = \alpha_{\tau} = A(E) & ; & \beta = \beta_{\tau'} = A(E).
\end{eqnarray}				
We expect the following alternative definition of conservation law to be equivalent to the one given: we say that the observable $A\in\mathcal{O}(\tau)$ is conserved under the evolution $V(\tau',\tau):\mathcal{U}_\tau\rightarrow\mathcal{U}_{\tau'}$ if
\begin{equation}
\label{6.3}		
d_{p_0,V}(\alpha,\alpha) = d_{p_0,V}(\beta,\beta)
\end{equation}				
for all $p_0\in \mathcal{S}(\tau_0)$ and all $E\in B(R)$. 

Let us verify the equivalence of these definitions in the history versions of traditional nonrelativistic quantum mechanics and classical mechanics.
\begin{enumerate}				
\item[(i)] Quantum Mechanics: Let $t_0 < t < t'$. We have $\mathcal{U}_{t_0}\approx\mathcal{U}_{t}\approx\mathcal{U}_{t'}=\mathcal{P}(\mathcal{H})$. With the initial state $\rho(t_0)=\rho_0$, $\alpha=\alpha_t\in\mathcal{P}(\mathcal{H})$ and $\beta=\beta_{t'}=\alpha_{t}$, we have (recalling eqs.(\ref{1.3}) and (\ref{1.5})) $C_\alpha = \alpha_t U(t,t_0)$, $C_\beta=\beta_{t'} U(t',t_0)$ and
\alphaeqn
\begin{eqnarray}
\label{6.4a}
d(\alpha,\alpha)=\mbox{Tr}(C_\alpha\rho_0 C^{\dagger}_\alpha)=\mbox{Tr}\left[\alpha_t U(t,t_0)\rho_0 U(t,t_0)^{-1}\right] \\
d(\beta,\beta) = \mbox{Tr}\left[\beta_{t'}U(t',t_0)\rho_0 U(t',t_0)^{-1}\right] 
\end{eqnarray}				
\reseteqn
The equality $d(\alpha,\alpha)=d(\beta,\beta)$ for arbitrary $\rho_0$ gives
\begin{equation}
\label{6.5}		
U(t',t_0)^{-1}\beta_{t'} U(t,t_0) = U(t,t_0)^{-1} \alpha_t U(t,t_0)
\end{equation}				
which implies (recalling that $\beta_t'=\alpha_t$)
\begin{equation}
\label{6.6}		
U(t',t)^{-1}\alpha_t U(t',t) = \alpha_t.
\end{equation}				
This equation represents the preservation of spectral projectors of the observable $A$ under Schr\"{o}dinger picture evolution. Consider, for example
\begin{displaymath}			
 \alpha_t = |\Psi_t\rangle\langle\Psi_t| \in \mathcal{U}_t
\end{displaymath}			
where $|\Psi_t\rangle$ is an eigenstate of $A$ corresponding to some eigenvalue $\lambda$. Under temporal evolution $|\Psi_t\rangle \rightarrow U(t',t)|\Psi_t\rangle$ and
\begin{equation}
\label{6.7}		
|\Psi_t\rangle\langle\Psi_t| \rightarrow U(t',t)|\Psi_t\rangle\langle\Psi_t|U(t',t)^{-1} \in \mathcal{U}_{t'}.
\end{equation}				
A general spectral projector of $A$ (which is a sum or integral of projectors of the form $|\Psi_t\rangle\langle\Psi_t|$) has the same transformation law under temporal evolution. The content of eq.(\ref{6.6}), therefore, is the same as that of eq.(\ref{6.1}) in the present case. The condition (\ref{6.3}), therefore, implies condition (\ref{6.1}). Conversely, given eq.(\ref{6.6}), the equality (\ref{6.3}) is easily obtained.
\item[(ii)] Classical Mechanics: Again, let $t_0 < t<t'$. We have $\mathcal{U}_{t_0}\approx\mathcal{U}_{t}\approx\mathcal{U}_{t'}= B(\mit\Gamma)$. With $p_0\in\mathcal{S}(t_0)$, $\alpha=\alpha_t=A(E) \in B(\mit\Gamma)$ and $\beta= \beta_{t'}=\alpha_t$ we have
\alphaeqn
\begin{eqnarray}
\label{6.8a}		
d(\alpha,\alpha) & = & \int_{\mit\Gamma} dp_0(\xi_0)\int_{\alpha_t} \delta\left[\xi-V(t,t_0)(\xi_0)\right] \nonumber \\
 & = & p_0\left[V(t,t_0)^{-1}(\alpha_t)\right]. 
\end{eqnarray}				
\begin{equation}
\label{6.8b}		
d(\beta,\beta)  =  p_0\left[V(t',t_0)^{-1}(\beta_{t'})\right].
\end{equation}				
\reseteqn
The equality $d(\alpha,\alpha)=d(\beta,\beta)$ for all $p_0$ gives 
\begin{equation}
\label{6.9}		
V(t,t_0)^{-1}(\alpha_t) = V(t',t_0)^{-1}(\beta_{t'}),
\end{equation}				
which (with $\alpha_t=\beta_{t'}=A(E)$ and assumed to be valid for all $E\in B(R)$) is easily seen to be equivalent to eq.(\ref{6.1}) in the present case. Conversely, given eq.(\ref{6.9}), eq.(\ref{6.3}) is easily deduced.
\end{enumerate}
						
Our definition of conservation law is somewhat different from the one given in [17] where, in the limited context of the traditional Hilbert space quantum mechanics (of a closed system [1,3]), conservation of an observable $A$ is defined in terms of vanishing probability of decoherent histories which involve projection operators corresponding to disjoint ranges of $A$ at two different times (making allowance for finite sequences of projection of other observables at intermediate times). Our definition, in contrast, is given (in a more general framework) in terms of equality of probabilities of single `time' histories at different `times' for all ranges of $A$ and for all initial states (see eq.(\ref{6.3})). The definition of [17] raises interesting questions about permitted projectors at intermediate times which are investigated there and some interesting results obtained.


\section{Connection between symmetries and conservation laws in Hilbert space-based theories}
\label{seven}

In certain forms of dynamics, one can obtain a general relation between continuous symmetries (\emph{i.e.} symmetries labelled by continuously varying parameters) of dynamics (generally expressed in terms of invariance of an action integral or of a Hamiltonian) and conservation laws. Examples are Lagrangian and Hamiltonian formulations of classical mechanics and Hilbert space quantum mechanics in the Heisenberg picture. The conserved quantities are infinitesimal generators of symmetry transformations (in the appropriate implementation of symmetry) interpreted as observables. In the present formalism, it appears difficult to see such a general connection between symmetries and conservation laws in the case of general $\mathcal{U}_\tau$'s. At least two conditions appear to be necessary to establish such a general connection: (i) an explicit expression for the decoherence functional $d_{p,V}$ (or at least some information about its dependence on the evolution maps $V(\tau',\tau)$) and (ii) presence of appropriate mathematical structure to ensure the identification of infinitesimal generator of a symmetry transformation as an observable. In [12], explicit expressions were given for the decoherence functional for two cases: in the Hilbert space-based theories and in classical mechanics. Of these, only the first class of theories satisfy the second requirement. (To satisfy the second requirement in classical mechanics, one will need to introduce smooth structures, and some aspects of the canonical formalism; we have not done that it this paper.) We shall prove below a theorem of the desired type in the Hilbert space-based theories; before doing that we quickly recall the relevant constructions.

This is the subclass of theories described in section \ref{two} in which we have, for each $\tau\in\mathcal{N}(\mathcal{T})$, a separable Hilbert space $\mathcal{H}_\tau$ and $\mathcal{U}_\tau = \mathcal{P}(\mathcal{H}_\tau)$, the family of projection operators in $\mathcal{H}_\tau$. Here $\mathcal{T}$ is a general space of temporal supports satisfying the condition of axioms $A_1$ and $A_2$ of section \ref{two}. For $\tau,\tau'\in\mathcal{N}(\mathcal{T})$ with $\tau\lhd\tau'$, the evolution from $\mathcal{U}_\tau$ to $\mathcal{U}_{\tau'}$ is, for the purpose at hand, more conveniently described by the linear map $K(\tau',\tau):\mathcal{H}_\tau\rightarrow\mathcal{H}_{\tau'}$ such that, for all triples $\tau,\tau',\tau''$ with $\tau\lhd\tau'\lhd\tau''$ and $\tau\lhd\tau''$, we have $K(\tau'',\tau')\cdot K(\tau',\tau) = K(\tau'',\tau)$. (When $\mathcal{H}_\tau$'s are naturally isomorphic and the maps $K(.,.)$ unitary, the transformation law $\alpha\rightarrow K(\tau',\tau)\alpha K(\tau',\tau)^\dagger$ of projectors gives the mappings $V(\tau,\tau') :\mathcal{P}(\mathcal{H}_\tau)\rightarrow\mathcal{P}(\mathcal{H}_{\tau'})$ which are logic isomorphisms.) It is also assumed that, for every pair $\tau,\tau'\in\mathcal{N}(\mathcal{T})$, there exists a $\tau''\in\mathcal{N}(\mathcal{T})$ such that $\tau\lhd\tau''$ and $\tau'\lhd\tau''$. (This is a new assumption not covered by the axioms.) Elements of $\tilde{\mathcal{U}}$ can be represented as homogeneous projectors of the form of eq.(\ref{4.4})
\begin{equation}
\label{7.1}		
\tilde\alpha = \alpha_{\tau_1}\otimes\alpha_{\tau_2}\otimes\cdots\otimes\alpha_{\tau_n}
\end{equation}				
where $\tau_1\lhd\tau_2\lhd\cdots\lhd\tau_n$ and $\alpha_{\tau_i}\in\mathcal{U}_{\tau_i} =\mathcal{P}(\mathcal{H}_{\tau_i})$, and those of $\Omega$ can be represented as inhomogeneous projectors (orthogonal sums of homogeneous projectors) as in the HPO formalism of section IV. It is adequate to give $d(\underline{\alpha},\underline{\beta})$ for homogeneous projectors $\underline{\alpha} = \{\tilde{\alpha}\}$, $\underline{\beta} = \{\tilde{\beta}\}$. Writing $\tilde\alpha$ for $\{\tilde{\alpha}\}$ and taking 
\begin{displaymath}		
\underline\beta = \beta_{\tau'_1}\otimes\cdots\otimes\beta_{\tau'_m}, 
\hspace{1cm} \tau'_1\lhd\cdots\tau'_m
\end{displaymath}		
we define (compare eq.(\ref{1.5}))
\begin{equation}
\label{7.2}		
d(\tilde\alpha,\tilde\beta)=N\mbox{Tr}\left[C'_{\alpha}\rho(\tau_0)C^{'\dagger}_\beta\right]		
\end{equation}				
where $\tau_0\lhd\tau_1$, $\tau_0\lhd\tau'_1$, $\rho(\tau_0)$ is a density operator on $\mathcal{H}_{\tau_0}$ and 
\begin{equation}
\label{7.3}		
C'_{\alpha}= K(\tau_f,\tau_n)\alpha_{\tau_n}K(\tau_n,\tau_{n-1})\cdots\alpha_{\tau_2}K(\tau_2,\tau_1)\alpha_{\tau_1}K(\tau_1,\tau_0)
\end{equation}				
with a similar expression for $C'_\beta$. Here $\tau_f$ is any element of $\mathcal{N}(\mathcal{T})$ satisfying the conditions $\tau_n\lhd\tau_f$ and $\tau'_m\lhd\tau_f$. Moreover $N^{-1}= \mbox{Tr}\left[A\rho(\tau_0) B\right]$ where $A$ and $B$ are the operators obtained from $C'_\alpha$ and $C'_\beta$ respectively by putting each of the $\alpha_{\tau_i}$ and $\beta'_{\tau_j}$ equal to the unit operator.

Since conservation laws can be defined only when all $\mathcal{U}_\tau$'s are isomorphic, we restrict ourselves to the subclass of Hilbert space-based theories in which all the $\mathcal{H}_\tau$'s are naturally isomorphic (and can, therefore be identified with a single Hilbert space $\mathcal{H}$) and the evolution maps are unitary. (This means that one has the usual Hilbert space-based quantum mechanics with unitary temporal evolution except that the time points $t,t',\cdots$ are replaced by the nuclear elements of a psg. Even this much generality, however, is worthwhile; results obtained will have validity in, for example, quantum field theories in a large class of space-times not admitting foliation in spacelike surfaces.)

We now proceed to obtain the desired relation between continuous symmetries and conservation laws.

Let $\Phi(\lambda) = (\Phi_1(\lambda),\Phi_2(\lambda),\Phi_3(\lambda))$ be a continuous 1-parameter symmetry of dynamics of a history system in the above mentioned class. (We shall be mainly concerned with $\Phi_2$.) According to some results obtained in section \ref{four} (see last-but-one para before eq.(\ref{4.20})), such a symmetry can be taken to be implemented unitarily in $\mathcal{H}$; let the corresponding infinitesimal generator be the self-adjoint operator $A$. The invariance condition (\ref{3.10}) of $V$ (with transformation law (\ref{3.7}) for $V(.,.)$) implies that the mappings $K(.,.)$ mentioned above commute with $A$ and therefore with the spectral projectors of $A$.

Now let $\tau_0\lhd\tau\lhd\tau'\lhd\tau_f$, $\tau_0\lhd\tau_f$ and $\tau\lhd\tau_f$ and consider the quantities $d(\alpha,\alpha)$ and $d(\beta,\beta)$ with $d(.,.)$ of eq.(\ref{7.2}) (with $N=1$) and $\alpha$, $\beta$ single `time' histories given by $\alpha_\tau = \beta_\tau = P$, a spectral projector of $A$. We have, in this case,
\begin{eqnarray}
\label{7.4}		
C'_\alpha & = & K(\tau_f,\tau) P K(\tau,\tau_0) \nonumber \\
 & = & K(\tau_f,\tau) K(\tau,\tau_0) P = K(\tau_f,\tau_0) P
\end{eqnarray}				
This gives
\begin{eqnarray}
\label{7.5}		
d(\alpha,\alpha) & = &\mbox{Tr}\left[C'_\alpha\rho(\tau_0)C^{'\dagger}_\alpha \right] \nonumber \\
 & = & \mbox{Tr}\left[K(\tau_f,\tau_0)P\rho(\tau_0)PK(\tau_f,\tau_0)^{\dagger}\right] \nonumber \\
 & = & \mbox{Tr}(P\rho(\tau_0)). 	
\end{eqnarray}				
Similarly, we have $d(\beta,\beta) = \mbox{Tr}(P\rho(\tau_0))$ giving the desired conservation law in the form of eq.(\ref{6.3}).


\section{Concluding remarks}
\label{eight}

The main inadequacy in the formalism of [12] which has seriously affected the present work as well is the absence of a concrete expression for the decoherence functional $d_{p,V}$. Construction of such a functional is an important problem which deserves serious effort at solution. There have been some attempts in literature [6,11,16,23] at construction of decoherence functional in various situations and at obtaining some general results about decoherence functionals [24,22,25]; these however, do not appear to be adequate to solve the above mentioned problem. 

The main achievement of the present paper is to show that even in the absence of such an expression, straightforward definition of symmetry can be given (which can be easily adapted to situations when a concrete expression for $d_{p,V}$ is available) and some interesting results obtained (both with and without decoherence functionals). An example of such a result obtained without a concrete decoherence functional is the formulation, in section \ref{three}, of a general criterion for physical equivalence of histories which covers the various notions of physical equivalence of histories considered by Gell-Mann and Hartle [14] as special cases. Examples of the results obtained using concrete decoherence functionals appeared in sections \ref{four}-\ref{seven}.


 
\end{document}